\documentclass[amsmath,prl,amssymb,floatfix,twocolumn]{revtex4}
\usepackage{amsthm,amsfonts,graphicx,verbatim,epsfig}

\usepackage{amssymb}

\newcommand{\Tr}{{\rm Tr}}
\newcommand{\D}{{\rm d}}

\newcommand{\curl}{{\bf \nabla}\times}
\def\req#1{(\ref{#1})}
\newcommand{\dd}{{\rm d}}

\newcommand{\R}{{\mathord{\mathbb R}}}
\newcommand{\ra}{\rangle}
\newcommand{\la}{\langle}
\newcommand{\tens}[1]{\overrightarrow{#1}\hspace{-4mm}\overleftarrow{#1}}

\newtheorem{thm}{Theorem}[section]
\newtheorem{lem}[thm]{Lemma}
\newtheorem{cor}[thm]{Corollary}
\newtheorem{prop}[thm]{Proposition}
\newtheorem{exa}[thm]{Example}
\newtheorem{rem}[thm]{Remark}
\newtheorem{definition}[thm]{Definition}

\begin{document}

\title{Casimir forces in a T operator approach}
\author{Oded Kenneth$^{1}$ and Israel Klich$^{2}$}%
\email{klich@caltech.edu} \affiliation{(1) Department of Physics, Technion, Haifa 32000
Israel \\
(2) Department of Physics, California Institute of
  Technology, MC 114-36 Pasadena, CA 91125}

\begin{abstract}
We explore the scattering approach to Casimir forces. Our main
tool is the description of Casimir energy in terms of transition
operators. The approach is valid for the scalar fields as well as
electromagnetic fields. We provide several equivalent derivations
of the formula presented in Kenneth and Klich [Phys. Rev. Lett.
97, 160401(2006)]. We study the convergence properties of the
formula and how to utilize it, together with scattering data to
compute the force. Next, we discuss the form of the the formula in
special cases such as the simplified form obtained when a single
object is placed next to a mirror. We illustrate the approach by
describing the force between scatterers in one dimension and three
dimensions, where we obtain the interaction energy between two
spherical bodies at all distances. We also consider the cases of
scalar Casimir effect between spherical bodies with different
radii as well as different dielectric functions.
\end{abstract}
\maketitle \vskip 2mm

\section{Introduction}
The Casimir force \cite{Casimir48} is one of the fundamental
predictions of quantum physics. It explores the interplay between
a quantum field and external "classical" like objects such as
boundary conditions, background dielectric bodies or space-time
metric. While the classical objects modify the behavior of the
field due to their presence, the field, in turn, acts on the
objects, typically by exerting forces. Much work has been devoted
to understanding the effect, as it appears in varied branches of
physics: from condensed matter (interaction between surfaces in
fluids) to gravitation and cosmology.

The first precise measurement of the effect by Lamoreaux
\cite{Lamoreaux97}, signaled a new age of Casimir force
measurements, and led to a revived interest in the theory behind
the effect. In recent years, the force between various objects
(such as two plates, plate and a sphere, corrugated plate and
sphere, etc (\cite{Lamoreaux97,MohideenRoy98,Bressi})) was
measured. Moreover the dependence on various properties of the
materials used , such as corrections due to finite conductivity
and temperature (\cite{Geyer,Pirozhenko Lambrecht Svetovoy06}), as
well as on geometry has been investigated. There is excellent
agreement between the experiments and the theoretical predictions,
which is being constantly improved. For an introduction to the
subject as well as reviews of progress see e.g.
\cite{BordagMohideenMostepanenko,Miltons Book,Millonis
Book,KardarGolestanian}.

Different variants (both material and geometric) of the force have
been proposed, discussed and motivated by pure theoretical
interest as well as by potential eventual application in
nano-mechanical structures \cite{Emig07,Ashourvan Miri
Golestanian}.

The original method used by Casimir, that of mode summation, has
led to a large body of work on the effect in simple geometries,
where the modes may be exactly computed. For more general cases
one has to use other available approaches such as the Green's
function approach or the path-integral approach. Significant
progress in utilizing these techniques numerically has been
reported lately \cite{Gies,RodriguezIbanescuIannuzzi}.

In the 1D case, scattering approach to Casimir physics has proved
very useful. Indeed, many of the calculations of Casimir
interaction between bodies are based on scattering theory, as the
photon spectrum in an open geometry is continuous and it's
description requires scattering.

In this paper, we explore a scattering approach to Casimir effect
in higher dimensions. The approach is based on analysis of a
determinant formula for Casimir interactions obtained in Ref
\cite{KennethKlich06}, and may be viewed as a generalization of
previous formulas, especially related to scattering, such as the
Lifshitz formula \cite{Lifshitz}, and the results of Balian and
Duplantier \cite{BalianDuplantier78}. Within this approach, the
Casimir energy is encoded in a determinant of the operator
$1-T_AG_0T_BG_0$ where $T_A,T_B$ are Lippmann-Schwinger $T$
operators associated with bodies $A$ and $B$ and $G_0$ is the
photons Green's function; we shall therefore refer to the formula
as the {\it TGTG} formula.

In \cite{KennethKlich06} it was shown how general results
regarding the direction of the force between bodies related by
reflection can be obtained from the $TGTG$ formula. For example,
the sign problem of interaction between two hemispheres was
resolved. This result was subsequently extended to a large class
of interacting fields possessing the "reflection positivity"
property \cite{Bachas} (See also \cite{ZNussinov}, where use is
made of reflection positivity arguments to infer attraction
between vortices and anti-vortices in a frustrated XY model). In
\cite{EmigGraham} an alternative derivation of the formula was
presented.

The paper is organized as follows. In section (II) we start with a
derivation of the determinant formula, as well as supply
alternative derivations in terms of Green's functions and the $T$
operator of a pair of perturbations. Section (III) illustrates how
one obtains the appropriate formula in the vector (electromagnetic
(EM)) case. Sections (IV) and (V) cover simplified cases: the
special case of a body placed next to a perfect mirror, and the
dilute limit, which deals with very weak dielectrics by expanding
round $\epsilon=1$.

We then proceed to show how the formula is to be applied in actual
calculations. We explain how the formula is to be used together
with partial wave expansions of the scattered states (sections
(VI) and (VII)). In 1D where only two modes (left and right
movers) exist at each $\omega$ this leads to a known closed form
formula for the Casimir energy in terms of reflection coefficients
(see, e.g. \cite{Kenneth99,Genet Lambrecht Renaud}).

In Section (VIII) we use spherical waves to obtain an explicit
expansion for the Casimir interaction between compact bodies. We
demonstrate this by computing the force between two spheres at all
distances, thereby generalizing the approach of
\cite{BulgacMagierskiWirzba} to spheres beyond Dirichlet boundary
conditions, and going beyond the proximity force approximation. We
also consider cases of spheres with unequal radii, as well as as
spheres with arbitrary dielectric function. In section (IX)
results are extended to the electromagnetic case. While this
manuscript was finalized we have learned that related results were
reported in \cite{EmigGraham} for the Casimir interaction between
spheres with equal radii, as well as an alternative derivation of
the determinant formula. 
Our results perfectly agree with those of
\cite{BulgacMagierskiWirzba} and \cite{EmigGraham}.

A number of appendices describe some technical details of the
calculations. Most notably, in appendix (B) we give additional
details about the mathematical validity of the formula, which were
not included in \cite{KennethKlich06}. These details, help
establish for the first time rigorously the validity of the
present approach to calculations of Casimir forces. In particular
we show that the formula is given in terms of $\log\det(1+A)$,
where $\Tr |A|<\infty$, and so mathematically well defined. (This
appendix is written in a "mathematical physics" style, and may be
skipped by readers not interested in these issues)

\section{The TGTG formula: Casimir
interaction as a regular determinant.} In this section, we explain
how the part of the free energy of a Gaussian theory that depends
on distance between bodies, and as such is responsible for the
Casimir force, may be expressed in terms of a regular determinant,
and discuss some of its properties. Some of the material covered
here appeared in the literature, however, as far as we know, the
final formula was never written in this general form; furthermore,
it's mathematical properties where not rigorously addressed
previously. We note, however that an elaborate and rigorous
analysis of a related problem involving impenetrable discs was
carried out in \cite{Wirzba97}.

We start by presenting the derivation of the determinant formula
\eqref{F for bulk} in the path integral approach
\cite{LiKardar,Kenneth99,FeinbergMannRevzen}. We first treat the
case of a scalar field  and explain later how the result is
extended to the EM field. Alternative derivations of Eq \eqref{F
for bulk} are elaborated in the following subsections.

The action of a real massless scalar field in the presence of
dielectrics can be written as
\begin{eqnarray}\label{scalar action}
S[\phi]={1\over 2}\int\dd^{d} {\bf r}\int {\dd\omega\over
2\pi}\phi_{\omega}^*(\nabla^2+\omega^2\epsilon({\bf
x},\omega))\phi_{\omega}
\end{eqnarray}
where $\phi_\omega^*=\phi_{-\omega}$, and $\epsilon(\omega,{\bf
x})=1+\chi({\bf x},\omega)$ is the dielectric function (we use
units $\hbar=c=1$). This action is the simplest action which
yields the scalar analog of the Maxwell equation
\begin{equation}
\curl \curl \overrightarrow{A}-{\omega^2\over
c^2}\epsilon(\omega,x) \overrightarrow{A}=0
\end{equation}
for the vector potential in the radiation gauge. Alternatively
this action can be derived by coupling a scalar field to an
auxiliary field living on the regions of space where $\epsilon\neq
1$, and then integrating out these fields, as done, e.g. in
\cite{FeinbergMannRevzen}.

Formally, the free energy of the system is obtained from the
partition function ${\cal Z}$ given by:
\begin{eqnarray}
  {\cal Z}=\int {\cal D}\,\,\phi e^{iS[\phi]}.
\end{eqnarray}
Performing the Gaussian integration, one finds that the change in
energy due to introduction of $\chi$ in the system is
\begin{eqnarray}\label{E(chi)-E(0) formal F}&
E_C=E_{\chi}-E_{\chi=0}=  \\ \nonumber & -i\int_0^{\infty}
{\dd\omega\over 2\pi}\log{\det}_{\Lambda}\left(1+\omega^2\chi({\bf
x},\omega) (\nabla^2+\omega^2+i0)^{-1}\right).
\end{eqnarray}

At this point, we encounter one of the main features of Casimir
physics - the need to properly isolate the physically relevant
part of the energy out of a formally ill defined expression. A
determinant (such as in Eq \eqref{E(chi)-E(0) formal F}) is
mathematically well defined only if it has the form $\det(1+A)$,
where $A$ is a "trace class" operator, i.e. $\sum_i
|\lambda_i|<\infty$ with $\lambda_i$ eigenvalues of $A$ (For
properties see appendix \ref{ccm} and \cite{Reed Simon
Scattering,Simon79}). If $A$ is not a trace class operator, one
may obtain different or infinite results for the determinant,
depending on the order in which the eigenvalues of $1+A$ are
multiplied. The expression above is not of the required form. To
see this note that $'A'$ in our case is given by:
\begin{eqnarray}
\omega^2\chi({\bf x},\omega) (\nabla^2+\omega^2+i0)^{-1}.
\end{eqnarray}
This is an operator of the form $g(x)f(\nabla)$. If such an
operator is "trace class" then it's trace is known to be given by
the Birman-Solomyak result \cite{Simon79}:
\begin{eqnarray}\label{Birman Solomyak}
 \Tr(g(x)f(i\nabla))= \int \D^3 x g(x)\int \D^3 k f(k)
\end{eqnarray}
in our case we have $\int \D^3 x\chi(x)<\infty$, however $\int
\D^3 k (-k^2+\omega^2+i0)^{-1}$ diverges, and is indicating that
the operator involved doesn't have a well defined trace
\footnote{This argument may be made precise by choosing a series
of functions for which the sum $\sum_{n=1}^N<\phi_n|A|\phi_n>$
diverges as $N\rightarrow\infty$}.

As such, the expression \eqref{E(chi)-E(0) formal F} only has
meaning when specifying physical cutoffs. Removing physical
cutoffs will leave us with an ill defined expression and so we
keep in mind cutoffs at high momenta in the notation
$\det_{\Lambda}$.

At high frequencies, $\chi(\omega,{\bf x})\rightarrow 0$ provides
a physical frequency cutoff. For ${\rm Re}\;\omega, {\rm Im}\;
\omega>0$ both $\chi(\omega)$ and $(\nabla^2+\omega^2+i0)^{-1}$ are
analytic, justifying Wick-rotation of the integration to the
imaginary axis $i\omega$ ending up with:
\begin{eqnarray}\label{analytical continuation of E}
E_{C}=\int_0^{\infty}{\dd\omega\over
2\pi}\log{\det}_{\Lambda}(1+\omega^2\chi({\bf x},i\omega) G_0({\bf
x},{\bf x}'))
\end{eqnarray}
Where ${G_0}({\bf x},{\bf x}')=\langle {\bf x}|{1\over
-\nabla^2+\omega^2}|{\bf x}'\rangle$. Restricting the operator
$(1+\omega^2\chi G_0)$ to the support of $\chi$ (more precisely to
$L^2(Supp(\chi))$) clearly does not affect its determinant. Note
that Eq.  \eqref{analytical continuation of E} is still ill
defined if one removes the cutoff, as can be immediately seen from
the argument based on Eq.  \eqref{Birman Solomyak}.

\begin{figure}
\includegraphics[scale=0.4]{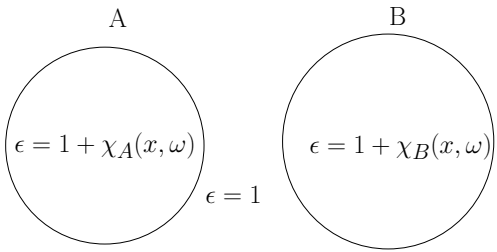}\caption{Bodies $A$ and $B$}
\label{Bodies}
\end{figure}

We now consider the case depicted in Fig. \ref{Bodies}, of two
bodies $A,B$ immersed in vacuum. $\chi$ is assumed nonzero only
inside the volumes of the two  dielectrics $A,B$ and we therefore
consider in the following $(1+\omega^2\chi G_0)$ as an operator on
${H_A\oplus H_B}\rightarrow{H_A\oplus H_B}$ where $H_A=L^2(A)$ and
$H_B=L^2(B)$ \footnote{Generalization to more then two dielectrics
is straightforward.}. It is then convenient to write
\begin{eqnarray}&
(1+\omega^2\chi G_0)\Big|_{H_A\oplus H_B}=\\ \nonumber &\left(
\begin{array}{cc}
1_A+\omega^2\chi_A {G_0}_{AA} & \omega^2\chi_A {G_0}_{AB} \\
\omega^2\chi_B {G_0}_{BA}& 1_B+\omega^2\chi_B {G_0}_{BB} \\
\end{array}
\right),
\end{eqnarray}
where ${G_0}_{\alpha\beta}$ is ${G_0}$ restricted to
$H_{\alpha}\rightarrow H_{\beta}$ (equivalently,
${G_0}_{\alpha\beta}=P_{\alpha}G_0P_{\beta}$, where $P_A=1\oplus
0$ and $P_B=0\oplus 1$, are projections on $H_A,H_B$
respectively).
 It turns out that the part of the energy that depends
 on mutual position of the bodies, and as such is responsible for the force, is a
well defined quantity, which is independent of the cutoffs. To see
this, we subtract contributions which do not depend on relative
positions of the bodies $A,B$:
\begin{eqnarray}\label{subtractions}
E_{C}=E_C(A\bigcup B)-E_C(A)-E_C(B)
\end{eqnarray}
As in Ref. \cite{FeinbergMannRevzen}, this amounts to subtracting
the diagonal contributions to the determinant, which are not
sensitive to the distance between the bodies, (i.e. only
contributes to their self energies). This yields
\begin{eqnarray}\label{manipulations for two bodies} &
E_{C}=\\ \nonumber & \int_0^{\infty}{\dd \omega\over 2\pi}
\Big\{\log{\det}_{\Lambda} \left(\begin{array}{cc}
1+\omega^2\chi_A {G_0}_{AA} & \omega^2\chi_A {G_0}_{AB} \\
\omega^2\chi_B {G_0}_{BA}& 1+\omega^2\chi_B {G_0}_{BB} \\
\end{array}\right)
\\ \nonumber &
-\log{\det}_{\Lambda}
\left(\begin{array}{cc} 1+\omega^2\chi_A {G_0}_{AA} & 0 \\
0& 1+\omega^2\chi_B {G_0}_{BB} \\ \end{array}\right) \Big\}\\
\nonumber & =\int_0^{\infty}{\dd \omega\over 2\pi}
(\log{\det}_{\Lambda}
\left(\begin{array}{cc} 1 & T_A{G_0}_{AB}\\
T_B{G_0}_{BA}& 1 \\  \end{array}\right),
\end{eqnarray}
where $T_{\alpha}={\omega^2\over 1+\omega^2\chi_{\alpha}
{G_0}_{{\alpha}{\alpha}} }\chi_{\alpha}$.
Finally, using the relation
\begin{eqnarray*}
\det\left(%
\begin{array}{cc}
  1 & X \\
  Y & 1 \\
\end{array}%
\right)=\det(1-YX),
\end{eqnarray*}
which holds for block matrices
we have:
\begin{eqnarray}\label{F for bulk}&
E_C(a)=\int_0^{\infty}{\dd \omega\over 2\pi}
\log\det(1-T_A{G_0}_{AB}T_B{G_0}_{BA}).
\end{eqnarray}
Henceforth, we refer to \eqref{F for bulk} as the {\it TGTG}
formula throughout the paper. Up to Wick rotation, the operators
$T_{\alpha}$ are exactly the $T$ operators appearing in the
Lippmann-Schwinger equation, as will be discussed in the next
subsections. The Wick rotation $T(\omega)\rightarrow T(i\omega)$ has
the effect of turning $T_{\alpha}$ into hermitian operators as
well as of avoiding potential singularities (which may occur at
real frequencies).

In Eq.  \eqref{F for bulk}, we disposed of the cutoff $\Lambda$ as
the expression is well defined in the continuum limit. In
practical terms this means that replacing the infinite dimensional
matrix of $1-T_A{G_0}_{AB}T_B{G_0}_{BA}$ by its its upper-left
$n\times n$ block with $n$ large enough and calculating the
resulting ordinary determinant, gives an arbitrarily good
approximation to a (finite) quantity which we call
$\det(1-T_A{G_0}_{AB}T_B{G_0}_{BA})$. This point is discussed in
detail in appendix \ref{mathp}, where we prove some mathematical
properties of the operators involved. The details are not
essential for understanding the applications of the formula, so a
reader not interested in mathematical rigor may skip them.

\subsection{Dirichlet and Neumann boundary conditions}

In many cases, and indeed in the original presentation by Casimir,
one is interested in sharp boundary conditions, such as Dirichlet
or Neumann. Sharp boundary conditions result in singular energy
density at the surface, as field modes are required to vanish for
all momentum scales. Typically, the local energy density diverges
as the inverse fourth power of the distance from the boundary
\cite{Candelas Deutsch}.

It is important to point out that the above considerations also
describe the conducting case with minor changes. Following
\cite{FeinbergMannRevzen}, assume conducting boundary conditions
are given over a surface $\Sigma$, parameterized by internal
coordinate $u$ and by the embedding in $\R^3$ given by ${\bf
x}(u)$. One may describe a simple metal by taking
$\chi(i\omega)={\omega_p^2\over 4\pi\omega^2}$ on $\Sigma$, and
letting $\Sigma$ have a thickness of a few skin depths
$l/\omega_p,\;l\sim{\cal O}(1)$, here $\omega_p$ is the plasma frequency
(proportional to the effective electron density in the metal). In
the limit of large $\omega_p$ one retains the same expression as
\eqref{F for bulk}, with the following substitutions:

\begin{eqnarray}\label{F for bulk Dirchlet Neumann}&
E_C(a)=\\ \nonumber & {1\over 2\pi}\int_0^{\infty}\D \omega
(\log\det(1- {\cal M}_{BA}{1\over 1+{\cal M}_A}{\cal
M}_{AB}{1\over 1+{\cal M}_B})
\end{eqnarray}
where in the Dirichlet case ${\cal M}$ is given by:
\begin{eqnarray}
M^{(D)}(u,u';\omega)=l\omega_p \sqrt{g(u)}G_0({\bf x}(u),
{\bf x}(u'))\sqrt{g(u')}
\end{eqnarray}
and acting on the surfaces $\Sigma$. Similarly Neumann boundary
conditions may be treated in the path integral method
 by taking \cite{Buscher Emig}:
\begin{eqnarray}&
M^{(N)}(u,u';\omega)=\\
\nonumber &\sqrt{g(u)}\sqrt{g(u')}\partial_{n(u)}
\partial_{n(u')}G_0({\bf x}(u),{\bf x}(u'))
\end{eqnarray}
in \eqref{F for bulk Dirchlet Neumann}. We remark, that rigorous
discussion of the formula in the Neumann case requires further
analysis which we did not pursue in this paper (see remarks after
Eq.  \eqref{vector action}).

\subsection{Derivation using Green's functions and T operators}
To make contact with Green's function approach we supply in this
section an alternative derivation of the TGTG formula. Most of the
derivation is standard and may be skipped by readers interested
only in new results. However, we point out that our approach where
the $T$ operator of combined scatterers are utilized seems new.
Here, we use the Green's function in order to express the density
of states (DOS) of a differential operator with background, and
then perform the mode summation by integrating over energies.

We briefly remind the reader some of the required material.
The standard discussion of this is usually done
in the context of non-relativistic quantum mechanics.
The retarded/advanced $G^{\pm}$ are then defined by:
\begin{eqnarray}
(E\pm is-{\cal H})G^{\pm}(E) =I
\end{eqnarray}
This equation should be understood as an operator
identity. If ${\cal H}$ is a differential operator, for example
${\cal H}=-\triangle$ then it is the operator form of the differential
equation:
\begin{eqnarray}\label{freeG}
(E\pm is+\triangle)G(x,x')=\delta(x-x')
\end{eqnarray}
Using the representation $\la
n|G^{\pm}(E)|n'\ra=\lim_{s\rightarrow0}{\delta_{nn'}\over E\pm
is-E_n}$, one then finds that the DOS is given by:
\begin{eqnarray}\label{rho0}
{1\over \pi}{\rm Im}\,\Tr G^{\pm}(E)=\mp\sum_n \delta(E-E_n)=\mp
\rho(E).
\end{eqnarray}
Noting the identity
\begin{eqnarray}
\partial_{E}\log[{E\pm is-E_n}]={1\over E\pm is-E_n},
\end{eqnarray}
one can rewrite this as
\begin{eqnarray}\label{rho1}
\rho(E)=\pm{1\over \pi}{\rm Im}\, \partial_{E}\Tr \log G^{\pm}(E).
\end{eqnarray}

We are more interested in the relativistic version of this.
(Indeed the Casimir force vanishes in the non-relativistic limit,
as the exchange of very massive virtual particles is suppressed.)
In the relativistic context the Feynman propagator $G$ is defined by
a similar formula to that of $G^+$:
\begin{eqnarray}
H(\omega^2+is)G =I
\end{eqnarray}
For example, the action (\ref{scalar action}) corresponds to
$H=-\triangle-\omega^2\epsilon$. In free space $\epsilon=1$ we
obtain the same equation as Eq.  (\ref{freeG}) apart from the
substitution $E\rightarrow\omega^2$. (There is also a not very
interesting conventional overall minus sign, which is the reason
some signs in the following equations are different from what the
reader may remember.) 
In the presence of nontrivial background (e.g. dielectric) the
$\omega$ dependence of $H$ can take quite an arbitrary form, a
fact that slightly complicates the derivation of the DOS. One may
take advantage of the relation
$${\rm Im}{F'(x \pm is)\over F(x \pm is)}=\mp\pi\sum_n \delta(x -x_n)$$
where $F(x )$ is any real function having simple zeroes at the
points $\{x_n\}$. Indeed, away from the zeroes $\{x_n\}$ the fact
that $F$ is real guarantees vanishing of the l.h.s while near the
zero $x_n$ we have ${\rm Im}{F'(x \pm is)\over F(x \pm is)}={\rm
Im}{F'(x_n)\over (x \pm is-x_n)F'(x_n)}= \mp{s\over (x
-x_n)^2+s^2}\rightarrow \mp \pi\delta(x -x_n)$. Generalizing the
relation from real functions $F(x)$ to hermitian operators
$H(\omega^2)$ \footnote{This may be justified by expanding in an
($\omega$-dependent) eigenstates basis of $H(\omega)$ and noting
that terms containing ${d\over d\omega}|n(\omega)\ra$ cancel.}
 allows writing
\begin{eqnarray}\label{rho2}
{\rm Im}\, \partial_{\omega}\Tr \log G(\omega)=-{\rm Im}\Tr H'(\omega)G(\omega)=
\pi\rho(\omega)
\end{eqnarray}
which is the obvious analog of Eq.  (\ref{rho1}). (Note however
that similar generalization of Eq.  (\ref{rho0}) would usually be
false.) In (\ref{rho2}) we implicitly assumed $\omega>0$ to avoid
an extra $sign(\omega)$ factor.

Now, assume that $G_0$ is known for $H_0$ and we add a
perturbation $V$, i.e.
\begin{eqnarray}
\left(H_0(\omega^2+is)+V(\omega^2+is)\right)G=I.
\end{eqnarray}
The change in DOS due to introduction of the potential $V$ is
formally:
\begin{eqnarray}
\Delta\rho={1\over \pi}{\rm Im}\, \partial_{\omega}\Tr \log GG_0^{-1}
\end{eqnarray}
We will be interested in the change in energy due to change in
the distance $a$ between two separated potentials $V_A$ and $V_B$, which make up $V$.
So we take $V=V_A+V_B$.

Thus,
\begin{eqnarray}\label{def of Delta rho}&
\partial_a\Delta\rho(\omega)=
{1\over\pi}{\rm Im}\,\partial_{\omega}\partial_a\Tr\log
(GG_0^{-1})= \\ \nonumber & {1\over\pi}{\rm
Im}\,\partial_{\omega}\partial_a\Tr\log
(I+G_0V)^{-1}.
\end{eqnarray}
Defining the $T$ matrix by
\begin{eqnarray}\label{Def of T} T=V(I+G_0V)^{-1},\end{eqnarray}
we may also write
\begin{eqnarray}
\partial_a\Delta\rho(\omega)={1\over\pi}{\rm
Im}\,\partial_{\omega}\partial_a\Tr\log (I-G_0 T).
\end{eqnarray}
Alternatively, formally writing $"\partial_a \det(V_A+V_B)=0"$
since $V_A,V_B$ act in different subspaces one can write that
\begin{eqnarray}\label{DOS using T}
\partial_a\Delta\rho(\omega)={1\over\pi}{\rm
Im}\,\partial_{\omega}\partial_a\Tr\log T.
\end{eqnarray}
this last formal expression, however, should be handled with care,
and so we avoid using it.

 The $T$ matrix satisfies:
\begin{eqnarray}\label{G=G_0+G_0TG_0}
G(\omega)=G_0(\omega)-G_0(\omega)T(\omega)G_0(\omega)
\end{eqnarray}
(here $\omega>0$ is actually $\omega+is$), and   frequently appears
in scattering theory. Also note $T=V-VGV$.


The operator $T$ appears in the Lippmann-Schwinger equation as
follows. Given a solution $\phi$ of the free equation, without a
potential $H_0(\omega)\phi=0$, one constructs a solution $\psi$ of
the eigenvalue equation $(H_0+V)\psi=0$ having the same incoming
part $\psi_{in}=\phi_{in}$. Formally, this is done by looking for
a solution of:
$$\psi=\phi-G_0V\psi,$$  which is the Lippmann-Schwinger equation.
It follows that
$\psi=(I+G_0V)^{-1}\phi
=(I-G_0T)\phi$, thus we may build a new solution $\psi$ from a
solution $\phi$ of the free equation. For example, choosing $\phi$
to be a plane wave solution, one obtains
\begin{eqnarray}
\psi_k=e^{ikx}-\int \D k' G_0(k')\la k'|T|k\ra e^{ik'x}.
\end{eqnarray}

Note that our relativistic normalization convention implies that
$T$ is related to the scattering matrix via $S=1-2\pi
i\delta(\omega^2-H_0)T$.

We now address the case of two potentials $V_A,V_B$. We assume for
simplicity that cutoffs are in place, and so work with the $T$
operators as matrices. We compute the joint transition matrix for
both perturbations $T_{A\bigcup B}$, and show that the part
independent on "self energy"  is exactly Eq.  \eqref{F for bulk}.

By using the formula \eqref{G=G_0+G_0TG_0} as $G_i=G_0-G_0T_iG_0$
(with $i=A,B$), together with  the
definition of $T$ \eqref{Def of T}, and straightforward algebraic
manipulations we obtain:
\begin{eqnarray}\label{1over G of two perturbations}&
{1\over 1+G_0 (V_A+V_B)}=(1-G_0T_A){1\over
1-G_0T_BG_0T_A}(1-G_0T_B)
\end{eqnarray}
and so the joint $T$ operator of a pair of perturbations may be
factored as
\begin{eqnarray}&
T_{A\bigcup B}=(V_A+V_B){1\over 1+G_0 (V_A+V_B)}=\\
\nonumber & (V_A+V_B)(1-G_0T_A){1\over 1-G_0T_BG_0T_A}(1-G_0T_B).
\end{eqnarray}
The important feature of this expression is the observation that
the only part of the expression which directly mixes between the
$A$ and $B$ is the factor $1-G_0T_BG_0T_A$. Indeed, plugging Eq.
\eqref{1over G of two perturbations} in Eq.  \eqref{def of Delta
rho} we see that the contribution of frequency $\omega$ to the
force is now given by:
\begin{eqnarray}& \nonumber
\partial_a\Delta\rho(\omega)=  {1\over\pi}{\rm
Im}\,\partial_{\omega}\partial_a\Tr\log (I+G_0(V_A+V_B))^{-1} =\\
\nonumber & {1\over\pi}{\rm
Im}\partial_{\omega}\partial_a[\log\det(1-G_0T_A)+\log\det(1-G_0T_B)-\\
\nonumber &
\log\det( 1-G_0T_AG_0T_B)]=\\
 & -{1\over\pi}{\rm Im}\partial_{\omega}\partial_a
\log\det( 1-G_0T_AG_0T_B)
\end{eqnarray}
leading again to our expression for the energy Eq.  \eqref{F for
bulk}.

Alternatively one may simply verify the correctness of Eq.
\eqref{F for bulk} by noting that:
\begin{eqnarray*}\label{1 over 1-G_0(V_1+V_2)}&
1-G_0T_AG_0T_B=1-G_0V_A{1\over1+G_0V_A}G_0V_B{1\over1+G_0V_B}=  \\
\nonumber & {1\over 1+G_0V_A} \left\{
(1+G_0V_A)(1+G_0V_B)-G_0V_AG_0V_B \right\}\\   \nonumber &\times
{1\over1+G_0V_B}= {1\over1+G_0V_A}[1+G_0(V_A+V_B)]{1\over1+G_0V_B}
\end{eqnarray*}
and using Eq.  \eqref{def of Delta rho}.

\section{The Electromagnetic Case}\label{Electromagnetic Field}
Here, we follow the approach of \cite{LifsitzPitaevskii}. The
statistical properties of the electromagnetic field in a medium
are described by the appropriate photonic Green's function. The
electromagnetic fields are derived from the electromagnetic
potentials $A^{\alpha}$, $\alpha=0,..,3$. (It is convenient to
work in the gauge $A^0=0$.)
 The retarded Green's function ${\cal D}_{ik}$
is defined by:
\begin{eqnarray}&
{\cal D}_{ik}(X_1,X_2)=\\ \nonumber & \Big{\{}\begin{array}{cl}
\la A_i(X_1)A_k(X_2)-A_k(X_2)A_i(X_1)\ra \,\,\,& t_1<t_2 \\  0  &
{\rm otherwise}
\end{array}
\end{eqnarray}
where $X_1,X_2$ are 4-vectors $X=(X^0,..X^3)$ and $k,i=1,..3$.
The angular brackets denote averaging with respect to the Gibbs
distribution.

The interaction of the electromagnetic field with a classical
current ${\bf J}$ put in the medium is given by $$ V=-{1\over
c}\int {\bf J}\cdot{\bf A}.$$

Kubo's formula allows us to treat this interaction within linear
response. By Kubo's formula the mean value $\overline{{\bf A_i}}$
in presence of a current ${\bf J}$ satisfies:
\begin{equation}\label{averageA}
\overline{{\bf A_i}}({\bf r})_{\omega}=-{1\over \hbar c}\int {\cal
D}^R_{ik}(\omega;{\bf r},{\bf r'}){\bf J}_{k}({\bf
r'})_{\omega}\D^3{\bf r'},
\end{equation}
where
\begin{equation}
{\cal D}_{ik}^R(\omega;{\bf r},{\bf r'})=\int_0^{\infty}e^{i\omega
t} {\cal D}^R_{ik}(t;{\bf r},{\bf r'})\D t
\end{equation}
The function ${\cal D}$ is sometimes referred to as the
generalized susceptibility of the system \cite{LifsitzPitaevskii}.

From Maxwell's equations it follows that in a medium with a given
permittivity tensor $\epsilon_{ij}$, permeability tensor
$\mu_{ij}$, and current ${\bf J}$, the vector potential $A_i$
satisfies:
\begin{equation}\label{MaxwellA}
(\curl(\mu^{-1}\curl)-{\omega^2\over c^2}\epsilon)\overline{{\bf
A}}={4\pi\over c}{\bf J}_{\omega}
\end{equation}
Substituting Eq.  \req{averageA} in Eq.  \req{MaxwellA}, we see
that ${\cal D}$ is a Green's function for the equation:
\begin{equation}\label{GreenEq} {\bf \nabla}\times\mu^{-1}{\bf
\nabla}\times {\cal D}-{\omega^2\over c^2}\epsilon {\cal
D}=-4\pi\hbar {\rm I}\delta({\bf r}-{\bf r}')
\end{equation}
where ${\rm I}$ is the three dimensional unit matrix. In the
following we shall work in units where $c=\hbar=1$.

The Green s function ${\cal D}$ is then used to obtain the well
known expression Eq.  (80.8) in Lifshitz and Pitaevskii
\cite{LifsitzPitaevskii} for the change in free energy due to
variation of the dielectric function $\epsilon$ at a temperature
$T$:
\begin{eqnarray}\label{Lifshitz Pitaevskii expr}\delta F=\delta
F_0+{1\over2}{T}\sum_{n=-\infty}^{\infty} \omega_n^2 \Tr ({\cal D}
\delta\epsilon).
\end{eqnarray}

Here $F_0$ is the free energy due to material properties not
related to long wavelength photon field, and $\omega_n=2\pi nT$
are Matsubara frequencies. ${\cal D}$ is the temperature Green's
function of the long wave photon field given by $ {\cal
D}(\vec{x},\vec{x'},i\omega)_{ij}=\la \vec{x}|{1\over
\nabla\times\nabla\times+\omega^2\epsilon(r,i|\omega|)}|\vec{x}'\ra_{ij}
$

Eq\eqref{Lifshitz Pitaevskii expr} may be written as $\delta
F=\delta F_0+\delta F_C$ where
\begin{eqnarray*}\label{free energy vector}&
F_C={1\over2}{T}\sum_{n=-\infty}^{\infty}
[\log{\det}_{\Lambda}(\nabla\times\nabla\times+\omega_n^2\epsilon(x,i\omega_n))\\
\nonumber &
-\log{\det}_{\Lambda}(\nabla\times\nabla\times+\omega_n^2)]\\
\nonumber & ={1\over2}{T}\sum_{n=-\infty}^{\infty}
\log{\det}_{\Lambda} (1+\omega_n^2\chi(x,i\omega_n){\cal
D}_0(i\omega_n)).
\end{eqnarray*}
Here $ {\cal D}_0(\vec{x},\vec{x'},i\omega_n)_{ij}=\la
\vec{x}|{1\over
\nabla\times\nabla\times+\omega_n^2}|\vec{x}'\ra_{ij}$. Note that
$F_C$ is exactly the same as \eqref{analytical continuation of E},
with the scalar propagator $G_0$ replaced by the vector propagator
${\cal D}_0$. For later reference we write here the explicit
expression for ${\cal D}_0$:
\begin{eqnarray}
{{\cal D}_0}_{ij}(k,i\omega)=-{4\pi\over
k^2+\omega^2}(\delta_{ij}+{k_ik_j\over\omega^2})
\end{eqnarray}
Thus, starting with this expression, one repeats Eq.
\eqref{subtractions} and Eq.  \eqref{manipulations for two bodies}
to get Eq.  \eqref{F for bulk}, replacing $G_0$ by ${{\cal D}_0}$
everywhere (including in the definition of the $T$ operators). The
analysis of the determinant now proceeds exactly as in the scalar
case.

Alternatively, the EM case may similarly be derived starting from
the functional determinant corresponding to an EM action analogous
to Eq.  \eqref{scalar action}. In the axial gauge ${\cal A}_0=0$
this action takes the form:
\begin{eqnarray}\label{vector action}
S={1\over 2}\int\dd^{3} {\bf r}\int {\dd\omega\over
2\pi}\vec{{\cal A}}_{\omega}^*(-\nabla\times\nabla\times
+\omega^2\epsilon({\bf x},\omega))\vec{{\cal A}}_{\omega}.
\end{eqnarray}

A  permeable body may similarly be described within our approach
by replacing the dielectric interaction term $\vec{\cal
A}\omega^2\chi(x)\vec{\cal A}$ in the Lagrangian by a magnetic
term: $\vec{\cal A}\nabla\times(1-{1\over\mu})\nabla\times
\vec{\cal A}$. One may then go on through our derivation using the
(differential) operator
$\nabla\times(1-{1\over\mu(x)})\nabla\times$ instead of
$-\omega^2\chi(x)$ everywhere. One major difference between the
two cases is worth noting:
whereas  the dielectric term is always described (after Wick
rotation) by a positive operator, the operator in the magnetic
term turns out to be negative (for $\mu>1$). This fact can be related to
the known Casimir electric-magnetic repulsion. Moreover, the ideal
$\mu\rightarrow\infty$ limit is seen to correspond to a Lagrangian
in which the term $(\nabla\times{\cal A})^2$ is missing (inside
the body) which makes a highly irregular lagrangian. Analogy with
a scalar field satisfying Neumann boundary conditions suggests that
this situation may be described by dropping the $(\nabla\phi)^2$
term inside the Neumann body. There are also other arguments in
favor of that approach\cite{kn}, however we did not bring  these
argument to a completely rigorous form.

\section{Dielectric in front of a mirror}\label{dielectric in
front of}

A somewhat simplified, but useful in practice, version of our
formula is obtained in the case of a body placed close to a
mirror. Consider the body $A$ to the left of a Dirichlet mirror $B$
located at $x_n=a/2$. It is well known (using the image method)
that the effect of the Dirichlet mirror is to replace the free
propagator $G_0$ by
\begin{eqnarray}
G_B({\bf x},{\bf x'})=G_0({\bf x},{\bf x'})-G_0({\bf x},J({\bf x'}))
\end{eqnarray}
where $J({\bf x}_{\|},x_{\perp})=({\bf x}_{\|},a-x_{\perp})$ denotes reflection
through the mirror plane.
This may be written as $G_B-G_0=-G_0{\cal J}$ where
${\cal J}$ is the operator defined by
${\cal J}\psi({\bf x})=\psi(J({\bf x}))$.
Noting the standard relation \eqref{G=G_0+G_0TG_0}\;
 $G_B=G_0-G_0T_BG_0$ between the Green
function in the presence of scatterer $B$ to its $T$ matrix one
concludes $G_0T_BG_0=G_0{\cal J}$ which when substituted in \eqref{F for
bulk} gives
\begin{eqnarray}\label{AD}
E_C(a)=\int_0^{\infty}{\dd \omega\over 2\pi}\log\det(1-{G_0}{\cal
J}T_A).
\end{eqnarray}

An alternative (though closely related) approach is to note that
by complete analogy to Eq.  \eqref{analytical continuation of E}
the energy it costs to place a body $A$ near a mirror $B$ is
$$E_{C}=\int_0^{\infty}{\dd\omega\over2\pi}\log{\det}_{\Lambda}(1+\omega^2\chi_A({\bf
x},i\omega) G_B({\bf x},{\bf x}')).$$ Subtracting the energy
$E_{C}=\int_0^{\infty}{\dd\omega\over2\pi}\log{\det}_{\Lambda}(1+\omega^2\chi_A({\bf
x},i\omega) G_0({\bf x},{\bf x}'))$ it cost to put $A$ in vacuum
then gives the Casimir interaction energy. Using the relation
\begin{eqnarray}&
(1+G_BV_A)/(1+G_0V_A)=\\ \nonumber & 1+(G_B-G_0)T_A=1-G_0{\cal
J}T_A
\end{eqnarray}
leads again to \eqref{AD}.

Yet another way of obtaining the same result is by substituting
$\chi_B=\lambda\delta(x_n-a/2)$ in the definition of $T_B$ and
doing the algebra. One then finds
\begin{eqnarray}& G_0T_BG_0=\\ \nonumber & \int{d^2k_\perp\over(2\pi)^2}e^{ik_\perp(x-x')_\perp}
{\lambda\omega^2\over 2q(\lambda\omega^2+
2q)}e^{-q|x_n|-q|x'_n|}\Big|_{q=\sqrt{\omega^2+k_\perp^2}},
\end{eqnarray}
 which
in the limit $\lambda\rightarrow\infty$ reduces, as expected, to
the expression $G_0{\cal J}$ obtained through the image method.

We now address the case of a Neumann mirror.  Note, that the Green
function in the presence of a Neumann mirror is $G=G_0+G_0{\cal J}$.
By repeating the arguments above we find that the Casimir
interaction between an object $A$ and a Neumann mirror is given by
a similar formula to \eqref{AD}, involving the determinant
$\det(1+ G_0{\cal J}T_A)$. We remark, that while the Dirichlet
mirror may be considered as the limit $\lambda\rightarrow\infty$
of a dielectric having e.g. $\chi_B=\lambda\delta(x_n-a/2)$ (or in
more realistic model $\chi_B=\lambda\theta(x_n-a/2)$) it is hard
to find a simple analog $\chi_B(x)$ that would lead in a similar
limit to a Neumann mirror. (See however remark at the end of the previous section).

A similar treatment is applicable in the more physically relevant
EM case. The boundary conditions $E_\|=0$ may be enforced by
requiring the vector potential to satisfy ${\cal J}A=-A$ where
${\cal J}$ is defined to act on vectors as
${\cal J}A(x)=(A_\|(J(x)),-A_\perp(J(x)))$ (Here $A_\|,A_\perp$ denote the
components of $A$ parallel and normal to the mirror surface. The
temporal component is considered as a parallel component though in
practice we usually choose a gauge where it vanishes.)
\newline The
EM Casimir interaction between a dielectric and a mirror is then
given by a formula similar to Eq.  \eqref{AD} with $G_0,{\cal J}$
replaced by the EM propagator ${\cal D}_0$ and the vectorial
${\cal J}$ defined above.

It is interesting to also consider an ideal permeable mirror
(having $\mu\rightarrow\infty,\epsilon=1$). This corresponds to
the boundary condition $B_\|=0$ which may be enforced by requiring
the vector potential to satisfy ${\cal J}A=+A$. Thus, the Casimir
interaction of body $A$ with such a mirror will be given by an
expression involving the determinant $\det(1+T_A{\cal D}_0{\cal
J})$.

\section{Dilute limit} In the following sections we consider strategies of using the
$TGTG$ formula in actual calculations. A particularly simple case
is when $\chi$ is small, which is commonly referred to as the
"dilute" case (and sometimes as "low contrast"). Here we briefly
sketch how to best use the formula in this limit. As shown in the appendix
(Theorem \ref{g<1}), one always have $||\sqrt{G}TGT\sqrt{G}||<1$, therefore we may expand the
$\log\det(1-..)$ expression \eqref{F for bulk} in powers:
\begin{eqnarray}\label{Dilute expansion}
  E_C=-\int{\dd\omega\over2\pi} \sum {1\over m}\Tr (T_AG_0T_BG_0)^m
\end{eqnarray}
In the dilute limit $\chi_{\alpha}<1$, so one may also substitute the
expansion
\begin{eqnarray}
T_\alpha=-\sum_{n=0}^\infty (-\omega^2\chi_\alpha
G_0)^n\omega^2\chi_\alpha
\end{eqnarray}
in Eq.  \eqref{Dilute expansion} and compute the involved
integrals to desired order. This expansion is the continuous
equivalent to summation of two body forces, and is equal to the
Born series appearing, for example, in Ref.
\cite{LifsitzPitaevskii}.

\section{Scattering Approach}
As remarked above, the operator
$T_AG_0T_BG_0$ appearing in our formula is closely related to
scattering data. The purpose of this section is to clarify this
relation and make it more explicit. In order to keep better touch
with conventions used in scattering theory, we usually avoid in
the following sections
using Wick rotation and thus we work in Lorentzian rather then
Euclidean space with real rather than imaginary frequency and with
the Feynman rather than the Euclidean propagator.

As mentioned above, the arguments of $G_0$ in Eq.  \eqref{F for
bulk} never coincide, implying that when $G_0(x_a,x_b)$ is
considered as a function of $x_b$ alone it is a solution of the
(homogeneous) free wave equation. Thus one may expand
$G_0(x_a,x_b)$ in the form $\sum {\cal
C}_{\alpha\beta}\phi_\alpha^*(x_a) \phi_\beta(x_b)$ where $\{
\phi_\alpha(x_a)\},\{\phi_\beta(x_b)\}$ are some sets of free wave
solutions of energy $\omega$. There is, of course, great freedom
in choosing the sets $\{ \phi_\alpha(x_a)\},\{\phi_\beta(x_b)\}$.
In practice one would choose these in a way that makes subsequent
calculations easier. Since we consider $T_AG_0T_BG_0$ as acting
only on the volume of object $A$, these considerations also apply
to the propagator on the right of this expression.

The Lippmann-Schwinger operator $T(\omega)$ is related to the
S-matrix by \footnote{Most standard textbooks discuss the
non-relativistic case and therefore include a factor
$\delta(\omega-\omega')$ instead of $\delta(\omega^2-\omega'^2)$.
Writing the delta function in terms of momentum the two cases
reduce to the same expression: $\delta(k^2-k'^2)$}
\begin{eqnarray}\label{NT}
S=1-2\pi i\delta(\omega^2-\omega'^2)T_\omega.
\end{eqnarray}
Therefore, $T(\omega)$ has the property that its matrix element
$\langle\alpha|T|\beta\rangle$ between a pair of free states
$\alpha,\beta$ having energy $\omega$ is equal to the
corresponding matrix element of the transition matrix. Since the
operator $T_B$ in $T_AG_0T_BG_0$ is sandwiched between a pair of
free Feynman propagators corresponding to energy $\omega$, we may
identify it with the corresponding transition matrix. Due to the
cyclicity of  the determinant $\det (1-T_AG_0T_BG_0)$ the same is
true of $T_A$.

Substituting the expansion $G_0(x_a,x_b)=\sum {\cal C}_{\alpha\beta}
\phi_\alpha^*(x_a)\phi_\beta(x_b)$ we arrive at
$$T_AG_0T_BG_0=\sum_{\alpha\alpha'\beta\beta'}T_A|\alpha\rangle {\cal C}_{\alpha\beta}
\langle\beta|T_B|\beta'\rangle {\cal
C}_{\alpha'\beta'}\langle\alpha'|$$ The Casimir interaction will
then be given explicitly by
\begin{eqnarray}\label{K}
E=\int_0^\infty{d\omega\over 2\pi}\log \det (1-K(i\omega)).
\end{eqnarray}
Here $K_{\alpha''\alpha'}= (T_A)_{\alpha''\alpha}{\cal
C}_{\alpha\beta}(T_B)_{\beta\beta'}{\cal C}_{\alpha'\beta'}$.

\section{Partial waves expansion}
In the following sections, we consider strategies of using the
representation \eqref{K} by restricting the $K$ matrix to a finite
subspace which gives the dominant contribution to the force.
Indeed, in many cases of interest only a few partial waves are
significantly scattered; the best example for this is when objects
are far apart, and from a large distance look point like. At this
limit one expects significant contribution only from $s$-wave
scattering. In the more general case, $K$ may be approximated by a
finite dimensional matrix corresponding to several partial waves.
In order to see how this works in practice we consider below a few
simple cases.
\subsection{One dimensional systems}
 A particularly simple case occurs when the system is
one-dimensional. Consider, e.g., a scalar field in 1D. All states
of energy $\omega$ are then spanned by two modes: left and right
movers $|L\ra,|R\ra={1\over\sqrt{2\pi}}e^{\pm i\omega x}$. Hence,
in this case, the determinant Eq (\ref{F for bulk}) can easily be
evaluated. To see how this is done, we write the Feynman
propagator explicitly as
\begin{eqnarray}
G_0=\int_{-\infty}^{\infty}{dk\over2\pi}{e^{ikx}\over\omega^2-k^2+i0}=-{i\over
2\omega}e^{i\omega|x|}
\end{eqnarray}
We consider a pair of scatterers $A,B$ such that $A$ is on the
left of $B$. This immediately implies that we have $x_a<x_b$ and
therefore
\begin{eqnarray}\label{G1d}
{G_0}_{BA}(x_b,x_a)=-{i\over 2\omega}e^{i\omega(x_b-x_a)}=
{-2i\pi\over2\omega}|R\rangle\langle R|
\end{eqnarray}
Similarly, we also have
${G_0}_{AB}={-2i\pi\over2\omega}|L\rangle\langle L|$. Using this
we see that the operator $K$ in Eq.  \eqref{K} turns into the
c-number
\begin{eqnarray}
K=({-2i\pi\over2\omega})^2\langle R|T_A|L \rangle\langle L|T_B|R\rangle=\tilde{r}_A(\omega)r_B(\omega).
\end{eqnarray}
Here $r_B$ ($\tilde{r}_A$) is the reflection coefficient for a
wave hitting scatterer $B$ from the left ($A$ from the right) to
be reflected back. Note that the normalization of $T$ implied by
Eq(\ref{NT}) is responsible to the cancelling of the factor
$-2i\pi\over2\omega$. (Had we used relativistic normalization for
$|L,R\ra$ the factor $2\omega$ would not have appeared.) We thus
conclude $$\det
(1-T_AG_0T_BG_0)=1-\tilde{r}_A(\omega)r_B(\omega).$$ The tilde on
$r_A$ serves to remind us that it is the reflection coefficient
from the right side of $A$.

We remark that $\tilde{r}_A(\omega)r_B(\omega)$ implicitly depends
on the distance between $A,B$ through the (phase) dependence of
$r_A,r_B$ on the scatterers locations. To make this explicit, note
that moving a scatterer a distance $a$ affects the reflection
coefficients as $r\rightarrow e^{-2ia\omega}r,\tilde{r}\rightarrow
e^{2ia\omega}\tilde{r}$.

Moving the scatterers a distance $a$ apart therefore results in
$$\det(1-T_AG_0T_BG_0)\rightarrow (1-e^{2ia\omega}\tilde{r}_A(\omega)r_B(\omega)).$$
Substituting in \eqref{K} we obtain the familiar formula for 1d
Casimir interaction between scatterers \cite{Kenneth99,Genet Lambrecht Renaud,Kampen}.

\subsection{Multi-component field in 1d}
The considerations used above for a single scalar field in one
dimension extend to a situation where
$\phi=(\phi_1,\phi_2,...\phi_n)$ is an $n$ component field. In
this case, the reflection coefficients $r_{A,B}$ turn into $n
\times n$ matrices and one finds $\det (1-T_AG_0T_BG_0)=\det
(1-\tilde{r}_A(\omega)r_B(\omega))$ where the determinant on the
right is of a usual $n \times n$ matrix.

\subsection{Plane wave expansion.}
In physical three dimensional space, there are many different
possible ways to expand the propagator $G_0(x_a,x_b)=\sum {\cal
C}_{\alpha\beta}\phi_\alpha^*(x_a) \phi_\beta(x_b)$ in terms of
free wave solutions $\{ \phi_\alpha(x_a)\},\{\phi_\beta(x_b)\}$.
In the next section we describe the expansion in spherical waves
(which is probably the most useful expansion), and we demonstrate
its use to calculating the Casimir force between compact object.
However, for the sake of simplicity we first describe here a plane
wave expansion which is the immediate generalization of Eq.
(\ref{G1d}). A simple heuristic way to arrive at this
generalization is to formally think of the field $\phi$ in three
dimensions as one dimensional field having infinitely many
components labelled by its transverse momenta. Indeed, such point
of view has been successfully used in describing transport in
quasi 1D conductors in mesoscopic physics, whereby each transverse
component corresponds to a scattering channel (see for example
Ref. \cite{Mello Stone}). This suggests splitting $\vec{k}$ into
its $z$-component $k_z$ and its transverse components
$k_\|=(k_x,k_y)$. The 3d propagator may then be written as:
$$G_0=-\int{d^2k_\|\over(2\pi)^2}
{ie^{i|z|k_z}e^{ik_\|x_\|}\over2k_z}\Big|_{k_z=\sqrt{\omega^2-k^2_\|+i0}}$$
Here $\sqrt{\omega^2-k^2_\|+i0}$ may be either real and positive
(for $\omega^2>k^2_\|$) or pure imaginary (for
$\omega^2<k^2_\|$) in which case the $i0$ prescription implies
that it must be chosen on the positive imaginary axis. Assuming
that $A$ is located to the left of $B$ along the $z$-axis it
follows that
\begin{eqnarray}\label{rep of TGTG}&
T_AG_0T_BG_0=\\   \nonumber &
\int{dk_xdk_ydq_xdq_y\over(2\pi)^4}
T_A|(q_x,q_y,-q_z)\rangle\times\\   \nonumber &
{1\over 2q_z}\langle(q_x,q_y,-q_z)|T_B|(k_x,k_y,k_z)
\rangle{1\over 2k_z}\langle(k_x,k_y,k_z)|,
\end{eqnarray}
where $q_z=\sqrt{\omega^2-q_x^2-q_y^2+i0}$ and
$k_z=\sqrt{\omega^2-k_x^2-k_y^2+i0}$.

When considering only the terms satisfying
$\omega^2>q_x^2+q_y^2,k_x^2+k_y^2$, Eq.  \eqref{rep of TGTG}
indeed looks like a straightforward generalization of the 1d
result. However, as this expression shows, to get the correct
result one must also include the contribution of evanescent waves
($q_\|^2>\omega^2$). Upon Wick rotation, however, the distinction
between ordinary and  evanescent waves disappears. It may also be
noted that (since in general $q_z\neq k_z$) the variation of the
$\langle(q_x,q_y,-q_z)|T_B|(k_x,k_y,k_z)\rangle$ matrix elements
upon moving $B$ along the $z$-axis is considerably more
complicated then in the 1d case.

The above representation may be helpful in problems where the
scatterers $A,B$ have exact or approximate planar geometry (e.g.
corrugated plates). Though the theorem guaranteeing finite trace
does not apply for infinite plates, one may show that dividing by
the plate area leads to finite result. We remark that actual
calculation of the determinant requires discretizing $k_\|$ which
corresponds to assuming large but finite plates. Alternatively one
may use Eq. (\ref{Dilute expansion}) with continuous $k_\|$.

\section{Spherical waves expansion}\label{sswe}

\begin{figure}
\includegraphics[scale=0.4]{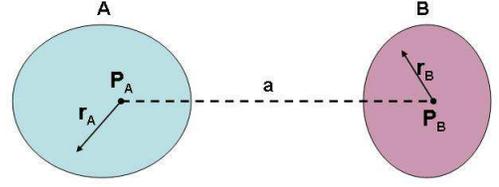}
\caption{Coordinate system used
for the partial wave approach} \label{coordinates}
\end{figure}

When describing interaction between two compact bodies, often it
is convenient to represent the transition matrices $T$ in a
spherical wave basis. To do so, we choose two points $P_A,P_B$
inside bodies $A,B$ respectively. We parameterize the points of
body $A$ by the radius vector $\vec{r}=\vec{r_A}$ measured from
the point $P_A$ and the points of $B$ by the radius vector
$\vec{r'}=\vec{r_B}$ measured from the point $P_B$ .
The vector connecting $P_A$ and $P_B$ will be denoted by $\vec{a}$
(Fig.\ref{coordinates}).
In the scalar case, the free spherical waves
centered at $P_A,P_B$ are given by
\begin{eqnarray} |(l m)_{A,B}\ra=\sqrt{2\omega^2\over\pi}
j_l(\omega r_{A,B})Y_{lm}(\hat{r}_{A,B}),
\end{eqnarray}
with the normalization $\langle\omega' l'm'|\omega
lm\rangle=\delta_{ll'}\delta_{mm'} \delta(\omega-\omega')$.

To use Eq. \eqref{K}, the scalar 3d Green function
$G_0=-{e^{i\omega r}\over4\pi r}$, is expanded in terms of the
spherical harmonic functions centered around $P_A$ and those
centered around $P_B$.
\begin{eqnarray}\label{YCY}
G_{\omega}=  \sum_{lm;l'm'}|(l m)_B\ra{{\cal
C}_{lm;l'm'}}\la(l'm')_A|
\end{eqnarray}
where
(See appendix \ref{AGB} for a proof of the following equations.)
\begin{eqnarray}\label{sg}&
{\cal C}_{lm;l'm'}(\omega)=\\ \nonumber & -{i\pi\over 2\omega} \sum_{l'',m''} C\left(\begin{array}{ccc} l & l' & l'' \\
m & m' & m''\\\end{array}\right) i^{l''+l'-l}h^{(1)}_{l''}(\omega a)
Y_{l''m''}(\hat{a}), \end{eqnarray}
$Y_{lm}$ are spherical harmonics, $j_l,h_l$ are spherical
Bessel and Hankel functions, and the coefficients
$$C\left(\begin{array}{ccc} l & l' & l'' \\
m & m' & m'' \\\end{array}\right)$$ have known expressions in
terms of the $3j$ symbol or as an integral of spherical functions:
\begin{eqnarray}\nonumber
C\left(\begin{array}{ccc} l & l' & l'' \\  m & m' & m''
\\\end{array}\right)= &
4\pi\int d\Omega Y_{lm}Y_{l'm'}^*Y_{l''m''}^*=\\
\nonumber= (-1)^{m} &
\sqrt{4\pi(2l+1)(2l'+1)(2l''+1)}\times\\
\label{C}
\left(\begin{array}{ccc} l & l' & l'' \\  0 & 0 & 0
\\\end{array}\right)
&\hspace{-16mm}
\left(\begin{array}{ccc} l & l' & l'' \\  m &
-m' & -m'' \\\end{array}\right)& \end{eqnarray}
In actual computations, it is often more convenient to use the
Wick-rotated expression. This may be expressed as ${\cal
C}_{lm;l'm'}(i\omega)=-{\pi\over2\omega}i^{l'-l}g_{lm;l'm'}$.
Where the coefficients
\begin{eqnarray}\label{sgW}&
g_{lm;l'm'}=\\ \nonumber & \sum_{l'',m''}
C\left(\begin{array}{ccc} l & l' & l'' \\ m & m' &
m''\\\end{array}\right) \sqrt{2\over\pi\omega
a}K_{l''+{1\over2}}(\omega a) Y_{l''m''}(\hat{a}), \end{eqnarray}
are real. Here $K_{l''+{1\over2}}$ are modified Bessel functions
of the second kind. Equations(\ref{sg},\ref{sgW}) may be somewhat
simplified by choosing the $z$-axis along $\hat{a}$.

The above expansion of $G_\omega$  allows expressing
$T_AG_0T_BG_0$ in terms of matrix elements $\langle
l'm'|T|lm\rangle$ of the transition matrices of the two
scatterers. The Casimir interaction may then be written as in Eq.
\eqref{K} where
\begin{eqnarray}&
K_{lm;l'm'}=\\ \nonumber & (-1)^{l_1+l_2}(T_A)_{lm;l_1m_1}{\cal
C}_{l_1m_1;l_2m_2} (T_B)_{l_2m_2;l_3m_3}{\cal C}_{l_3m_3;l'm'}.
\end{eqnarray}
Here ${\cal C}_{lm;l'm'}$ is given by Eq. \eqref{sg} or Eq.
(\ref{sgW}), summation over $l_1,m_1,l_2,m_2,l_3,m_3$ is implied
and we note that the extra sign resulted from ${\cal
C}_{lm;l'm'}(-\hat{a})\equiv(-1)^{l+l'} {\cal
C}_{lm;l'm'}(\hat{a})={\cal C}_{l'm';lm}(\hat{a})$.

If we assume that only waves having $l\leq l_0$ are significantly
scattered then $K$ will turn into a finite $(l_0+1)^2\times
(l_0+1)^2$ matrix (since the dimension of the subspace $l\leq l_0$
is $\sum_{l=0}^{l_0} (2l+1)=(l_0+1)^2$). We stress that this
argument {\it does not} require us to assume spherical symmetry of
the scatterers.

When $A,B$ are very far apart, the interaction between them is
governed by waves of very low frequency and therefore also low
$l$. At this limit the leading contribution comes from the
$s$-wave scattering transition matrix element $\langle
l=0|T_{A,B}|l=0\rangle\simeq 2\omega^2\lambda_{A,B}/\pi$, where
$\lambda$ is the scattering length.

The matrix $K$ then reduces to the scalar
$K=-\omega^2\lambda_A\lambda_B\left( h_0^{(1)}(\omega
a)\right)^2=4\pi{\lambda_A\lambda_B\over a^2}e^{2ia\omega}$.
Doing the integral (\ref{K}) one arrives at
$$E_C=-{\lambda_A\lambda_B\over a^3}.$$ This limit corresponds to
the scalar version of the well known Casimir-Polder interaction.
Our formalism, however, allows calculating corrections to it up to
any desirable finite order in $1\over a$. 
For example for two Dirichlet spheres of radii $R_1,R_2$
at  distance $a$ between their centers the expansion gives:
\begin{eqnarray}
E=&-{R_1R_2\over4\pi a^3}-{R_1R_2(R_1+R_2)\over 8\pi a^4}\\ \nonumber
&-{R_1R_2(34R_1^2+9R_1R_2+34R_2^2)\over48\pi a^5}\\   \nonumber
&-{R_1R_2(R_1+R_2)(2R_1^2+21R_1R_2+2R_2^2)\over36\pi a^6}+...
\end{eqnarray}

\subsection{Spherical scatterers}
Significant simplification is possible whenever $A,B$ have spherical
symmetry. First, the $T$-matrices are diagonal in angular momentum
basis and so may be expressed as
$$\langle l'm'|T_{A,B}|lm\rangle=\delta_{ll'}\delta_{mm'}
{2i\omega\over 2\pi}(e^{2i\delta_l^{A,B}(\omega)}-1).$$
where the normalization factor ${2i\omega\over 2\pi}$ follows from Eq(\ref{NT}).
A second consequence is that rotation around $\hat{a}$
(which from now on we take as coinciding with $\hat{z}$ axis)
is a symmetry of the whole system. The determinant therefore
factors as a product of terms corresponding to different values
of the azimuthal number $m$. The energy turns into a sum of the
corresponding terms
$$E=\sum_m\int{d\omega\over2\pi}\log\det(1-K^{(m)}(i\omega)).$$

The matrices $\{K^{(m)}_{ll'}\}_{ll'=|m|}^\infty$ defined for each
$m\in\mathbb{Z}$ (actually $K^{(-m)}=K^{(m)}$) are infinite
dimensional but may be approximated in numerical calculations by
finite matrices corresponding to $ l,l'\leq$ some $l_0$. The
operator $K^{(m)}$ may be written explicitly as
$$K_{ll'}^{(m)}=\sum_j g_{lj}^{(m)}t^{(A)}_jg_{jl'}^{(m)}t^{(B)}_{l'},$$
where we used the notation:
$$t_j={1\over2}(-1)^j(e^{2i\delta_j}-1),$$
\begin{eqnarray}&
g_{l_1,l_2}^{(m)}(i\omega)=(-1)^{m+1} \sqrt{(2l_1+1)(2l_2+1)}\\
\nonumber & \sum_{l}(2l+1) \sqrt{2\over\pi
a\omega}K_{l+{1\over2}}(a\omega) \left(\begin{array}{ccc} l_2 &
l_1 & l  \\  0 & 0 & 0
\\\end{array}\right) \left(\begin{array}{rrr} l_2 & l_1 & l \\  m &
-m & 0 \\\end{array}\right)
\end{eqnarray}
Note that both $g_{l_1,l_2}^{(m)}(i\omega)$ and $t_j(i\omega)$ are real.

\subsection{Dirichlet Spheres}
The simplest example for which the above may be applied is the
interaction of two hard (Dirichlet) spheres. The $T$-matrix
elements are well known in this case, and are given by
$t_j(\omega)=(-1)^{j+1}{j_{j}(\omega R)\over h^{(1)}_{j}(\omega
R)}$ which translates to
\begin{eqnarray}
t_j(i\omega)={\pi\over2}{I_{j+{1\over2}}
(\omega R)\over K_{j+{1\over2}}(\omega R)}
\end{eqnarray}
on the imaginary frequency line. Here $R$ is the sphere's radius
and $I_{j+{1\over2}}$ is a modified Bessel function of the first
kind.

In the special case where the two spheres have equal radii $R_1=R_2$
an extra simplification occurs. One can then write
$K=\tilde{K}^2,\tilde{K}_{ll'}=g_{ll'}t_{l'}$,
which imply $\log\det(1-K)=\log\det(1+\tilde{K})+\log\det(1-\tilde{K})$.
The numerical calculation of the two determinants $\det(1\pm\tilde{K})$ is then
somewhat easier then direct calculation of $\det(1-K)$.
Moreover comparison with section \ref{dielectric in front of}
shows that the two determinants  $\det(1\pm\tilde{K})$
(actually with $\tilde{K}_{ll'}=(-1)^m g_{ll'}t_{l'}$)
correspond to the Casimir
interaction energies $E_{D,N}$ of a single hard sphere and a
Dirichlet/Neumann mirror a distance $a/2$ away.
The symmetric two hard sphere system then has the energy $E_S=E_D+E_N$.
(One may also understand this in terms of decomposition into even and odd modes.)

We have done the calculation including partial waves of $l\leq
l_0$ for different values of $l_0$ and considered the $l_0$
dependence of the results as a test for convergence. Most
calculations included modes up to $l_0=10$, but for small values
of sphere separation $a$, we used larger $l_0$ even up to $l_0=72$
for $a/R=2.1$. Since we expected the error to behave roughly as
$E_c-E(l_0)\sim O(e^{-cl_0})$ we tried to fit the results with
this assumed asymptotics. The numbers suggests that in both the
sphere-sphere and in the sphere-plate cases we have
$c\sim2\log(1+d/R)$
where $d$ is the distance between the two objects
(i.e. $d=a-2R$ for $E_S$ and $d=(a-2R)/2$ for $E_D,E_N$).
Table(\ref{cl}) below shows the value of the constant $c$ for identical
spheres as a function of their distance as well as the value of
$l_0$ at which the error dropped to within $1\%$ of the exact
result. It should be remarked that the estimate for $c$ is a bit
crude since our numerics is consistent with $c$ being a slowly
growing function of $l_0$ (which might be due to sub-leading
asymptotics). By matching our results with the assumed
asymptotics, one can obtain a corrected estimate for $E_C$.
Comparison of this estimate with results obtained by increasing
$l_0$ gave good agreement.

\begin{eqnarray}
\begin{tabular}{|l|l|l|}
\hline
$a/R$&$ c $ &$L(1\%)$\\
\hline
2    &  0      & $\infty$ \\
2.1  &  0.18   &   31    \\
2.2  &  0.34   &   16    \\
2.35 &  0.57   &    9    \\
2.5  &  0.78   &    7    \\
2.75 &  1.06   &    5    \\
3    &  1.33   &  3-4     \\
3.5  &  1.78   &  2-3     \\
4    &  2.14   &    2    \\
5    &  2.75   &    1    \\
7    &  3.44   &    1    \\
\hline
\end{tabular}
\label{cl}
\end{eqnarray}

\begin{figure}
(a)\includegraphics[scale=0.8]{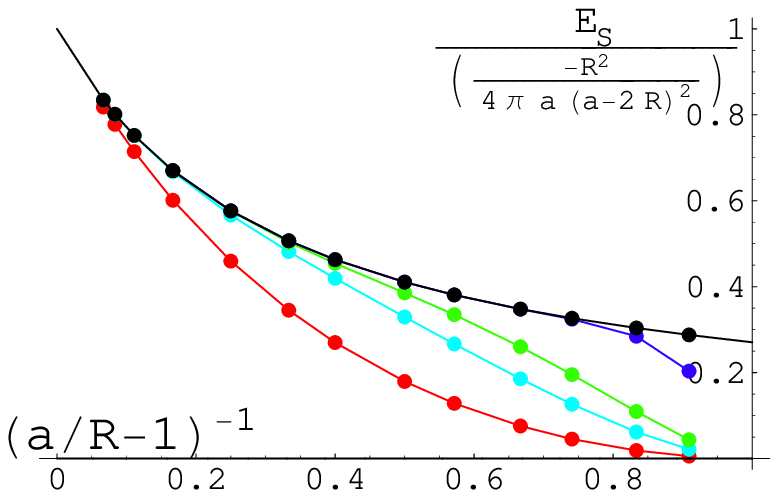}\vskip 0.5cm
(b)\includegraphics[scale=0.8]{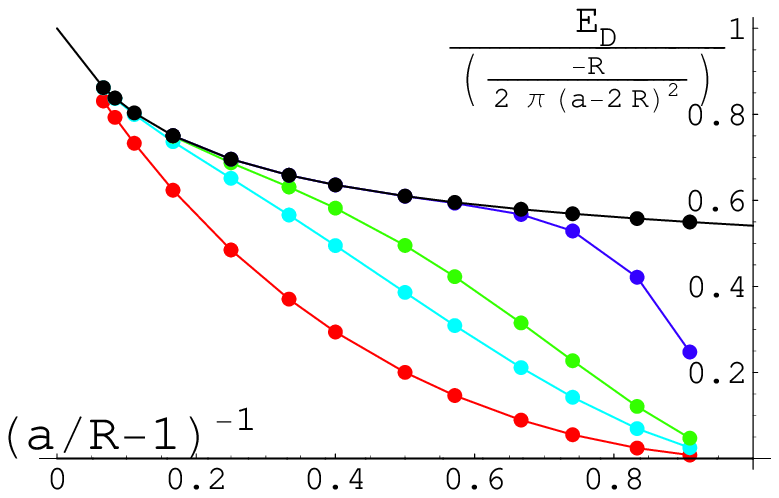}\vskip 0.5cm
(c)\includegraphics[scale=0.8]{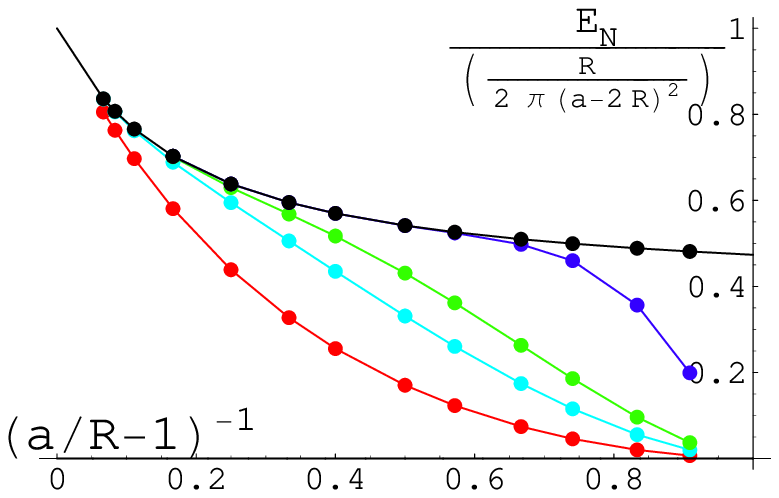} \caption{The calculated
Casimir energy of: (a) Two Dirichlet spheres of radius $R$ at
distance $a$ between their centers. (b,c) A Dirichlet sphere of
radius $R$ whose center is at a distance $a/2$ from a
Dirichlet/Neumann mirror. 
The graphs show $E/E_0$ as a function of $(a/R-1)^{-1}$ where
$E_0$ is the large distance asymptotic expression of it.
Specifically: (a) $E_0^S=-{R^2\over4\pi(a-2R)^2a}$, (b,c)
$E_0^{D,N}=\mp{R\over2\pi(a-2R)^2}$. At short distances $E/E_0$
approach the PFA prediction (a)${\pi^4\over360}\sim0.27$,
(b)${\pi^4\over180}\sim0.54$, (c)${7\pi^4\over1440}\sim0.47$. 
The black curve shows the calculated exact result for
$l_0\rightarrow\infty$. We extrapolated it to $a=2R$ and
$a=\infty$ using the known asymptotics. The colored graphs show
the result of including partial waves of $l\leq l_0$ where:
$l_0=0$ (red), $l_0=1$ (sky blue),
$l_0=2$ (green), $l_0=10$ (blue).}
\label{EE}
\end{figure}

The following table (\ref{edns}) and Fig.\ref{EE} show the results for the
Casimir energy itself (measured in units of $\hbar c\over R$).
$E_D$ denotes the energy of Dirichlet-mirror+(Dirichlet)sphere
system, $E_N$ denotes the energy of
Neumann-mirror+(Dirichlet)sphere system, $E_S$ denotes the energy
of the symmetric two hard sphere configuration (having
$E_S=E_D+E_N$). The result for $E_D$ are in perfect agreement with
a similar calculation done in Ref. \cite{BulgacMagierskiWirzba}.

\begin{eqnarray}
\begin{tabular}{|l|l|l|l|}
\hline
$a/R$&$ E_D$ &$E_N$&$E_S$\\
\hline
2.1& -8.75& 7.66& -1.0939\\
2.2& -2.2129& 1.9382& -0.27477\\
2.35& -0.739& 0.6488& -0.0902822\\
2.5& -0.3688 & 0.3245 & -0.0443005\\
2.75& -0.1679& 0.1483& -0.0195891\\
3& -0.09703& 0.08613& -0.0108937\\
3.5& -0.044981& 0.0403034& -0.00467768\\
4& -0.0261973& 0.0236767& -0.00252067\\
5& -0.0123048& 0.0112853& -0.00101948\\
7& -0.00477708& 0.00447243& -0.000304649\\
10& -0.00199796& 0.00190445& -0.0000935083\\
13& -0.0011022& 0.00106165& -0.0000405423\\
16& -0.000700129& 0.000678957& -0.000021172\\
\hline
\end{tabular}
\label{edns}
\end{eqnarray}



It may be remarked that $E_D,E_N$ correspond to sphere-mirror
distance, which is half the sphere-sphere distance in the
corresponding calculation of $E_S$. This fact is responsible among
other things to slower convergence in the calculation of $E_D,E_N$
and hence to a smaller number of calculated significant digits
compared with $E_S$.


We would like to mention two points regarding the actual implementation
of the numerical calculation. (Our earlier numerical attempts failed because we
were not fully aware of these points.)

The expressions of $g_{ll'}(i\omega),t_l(i\omega)$ may attain at small
$\omega$'s very large/small values respectively in such a way that only
their product remain finite. At large $\omega$'s similar phenomena occur
with $t_l(i\omega)$ large and  $g_{ll'}(i\omega)$ small. Thus to
avoid computer overflow it is much better to ``renormalize'' these two
quantities redefining $\tilde{g}_{ll'}=z_lz_{l'}g_{ll'},
\tilde{t}_l=t_l/z_l^2$ with $z_l\sim(R\omega)^{l+1/2}e^{R\omega}$.

A second important point is that one should make sure that the
computer program doing the calculation does not use the expansion
of $I_{l+1/2}(x)$ in terms of elementary functions. In MATHEMATICA
(which we used) this expansion is an automatic default whenever
the index of the Bessel function is half integer. However this
expansion is known to be numerically unstable (except for very
small $l$) and using it would lead to errors.

The general formula works well for $R_1\neq R_2$. For example
taking $R_1=R_0,R_2=2R_0$ and measuring $E$ in units of $\hbar c\over R_0$
we found

\begin{eqnarray}&
\nonumber \text{\it Interaction energy between a sphere of}  \\&
\nonumber \text{\it radius $R_0$ and a sphere of radius $2R_0$}        \\&
\begin{tabular}{|l|l|}
 \hline
$a/R_0$&$ E$\\
\hline
3.1 & -1.4554\\
3.2 & -0.367535\\
3.3 & -0.164591\\
3.4 & -0.0931057\\
3.5 & -0.0598295\\
3.67 & -0.0334525\\
3.83 & -0.021821\\
4 & -0.01501511\\
5 & -0.00362365\\
6 & -0.00151965948\\
8 & -0.00047970126\\
10 & -0.00021536976316\\
14 & -0.0000696380241\\
18 & -0.00003103693506\\
22 & -0.0000164921322\\
\hline
\end{tabular}
\label{r1=2r2}
\end{eqnarray}

\subsection{Dielectric Spheres}
\begin{figure}[tbh]
\includegraphics[scale=0.8]{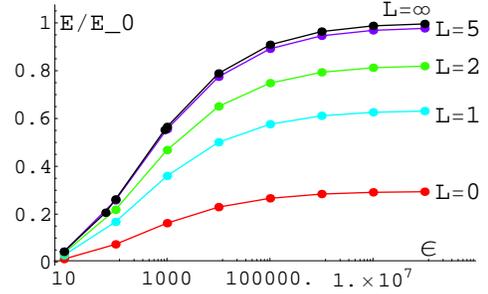}
\caption{The calculated Casimir energy of two (scalar) Dielectric spheres
of radii $R_2=2R_1$ at centers distance $a=4R_1$ depicted as a function of
their dielectric constant $\epsilon_1=\epsilon_2$. The energy $E$ was
normalized by the Dirichlet spheres result $E_0$ so that at
$\epsilon\rightarrow\infty$ we obtain $E/E_0=1$.}\label{dd}
\end{figure}

The formula also works well for finite dielectric constant. For
example the numerical results for $R_1=R_0;R_2=2R_0;a=4R_0$ as a
function of $\epsilon_1=\epsilon_2$ are given by the following
table(\ref{ee}) and Fig. \ref{dd}.
\begin{eqnarray}
\begin{tabular}{|l|l|}
\hline
$\epsilon$ & $ E$\\
\hline
64 &   -0.003092\\
100 &  -0.003927\\
900  &  -0.00829\\
$10^3$ & -0.0084835\\
$10^4$  & -0.01184\\
$10^5$  & -0.01364\\
$10^6$ &   -0.01447\\
$10^7$ &   -0.01483\\
$10^8$  &  -0.01495\\
$\infty$ & -0.01501511\\
\hline
\end{tabular}
\label{ee}
\end{eqnarray}

The calculation may easily be repeated for
any given $R_1,R_2,\epsilon_1,\epsilon_2,a$.

%
%

\section{Electromagnetic field}
To extend the ideas of the previous section from the scalar to the EM case
one needs to present the EM propagator in a form  analogous to eqs. \eqref{YCY}-\eqref{C}.
The required representation of the EM propagator (derived in appendix \ref{AGB}) is

\begin{eqnarray}\label{D_0expansion}
\tens{\cal D}_0=|(jm\alpha)_B\ra{\cal C}_{jm\alpha;j'm'\alpha'}\la (j'm'\alpha')_A|
\end{eqnarray}
where $\alpha,\alpha'$ can take the two values $0,1$ corresponding to the {\bf TE}
(magnetic multipole) or {\bf TM} (electric multipole) modes respectively.
The $\cal C$ coefficients are given by
\begin{eqnarray}\label{celm}&
{\cal C}_{jm\alpha;j'm'\alpha'}=\\
\nonumber & -{2i\pi^2\over\omega}\sum_{l''m''}
i^{l''+j'-j+\alpha-\alpha'}h^{(1)}_{l''}(\omega
a)Y_{l''m''}(\hat{a}) \\ \nonumber & \times \int d\Omega
Y_{l''m''}^*(\vec{Y}_{jm}^{(\alpha)}\cdot\vec{Y}_{j'm'}^{(\alpha')*})
\end{eqnarray}
where $\vec{Y}_{jm}^{(\alpha)}$ may be defined in terms of vectorial spherical harmonics as
\begin{eqnarray}&
\vec{Y}_{jm}^{(0)}=\vec{Y}_{jjm},\;\\ &
\vec{Y}_{jm}^{(1)}=\sqrt{j+1\over2j+1}\vec{Y}_{j,j-1,m}+\sqrt{j\over2j+1}\vec{Y}_{j,j+1,m}
\label{Yvec}
\end{eqnarray}
These functions satisfy $i\vec{Y}_{jm}^{(1)}=\hat{r}\times \vec{Y}_{jm}^{(0)},\;
i\vec{Y}_{jm}^{(0)}=\hat{r}\times \vec{Y}_{jm}^{(1)}$.


After Wick rotating, we obtain ${\cal
C}_{jm\alpha;j'm'\alpha'}(i\omega)=
i^{j'-j+\alpha-\alpha'}{\pi\over2\omega}g_{jm\alpha;j'm'\alpha'}$
where the coefficients
\begin{eqnarray}\label{icelm}
& g_{jm\alpha;j'm'\alpha'}(i\omega)=\sqrt{32\pi\over\omega a}\\
\nonumber & \times \sum_{l''m''} K_{l''+{1\over2}}(\omega
a)Y_{l''m''}(\hat{a})\int d\Omega
Y_{l''m''}(\vec{Y}_{jm}^{(\alpha)*}\cdot\vec{Y}_{j'm'}^{(\alpha')})
\end{eqnarray}
are real.



The integrals $\int d\Omega Y_{j_3m_3}(\vec{Y}_{j_1m_1}^{(\alpha)*}\cdot\vec{Y}_{j_2m_2}^{(\alpha')})$
appearing in (\ref{celm},\ref{icelm})
can be expressed explicitly in terms of 3j-symbols as follows.
For $\alpha=\alpha'$ it is given by
\begin{eqnarray}\label{3j0}&
{j_1(j_1+1)+j_2(j_2+1)-j_3(j_3+1)\over2\sqrt{j_1(j_1+1)j_2(j_2+1)}}
\sqrt{(2j_1+1)(2j_2+1)(2j_3+1)\over4\pi}\\ \nonumber  &\times
\left(\begin{array}{ccc} j_1 & j_2 & j_3 \\  0 & 0 & 0
\\\end{array}\right) \left(\begin{array}{ccc} j_1 & j_2 & j_3 \\
-m_1 & m_2 & m_3 \\\end{array}\right)\end{eqnarray}
(which vanishes unless $j_1+j_2+j_3\equiv0 mod2$.)
For $\alpha\neq\alpha'$ the integral is nonzero only provided
$j_1+j_2+j_3\equiv1 mod2$ in which case it is given by
\begin{eqnarray}\label{3j1}& (-1)^{m_1}\sqrt{(2j_1+1)(2j_2+1)(2j_3+1)\over4\pi}\\
\nonumber & \times\left(\begin{array}{ccc} j_1 & j_2 & j_3 \\  1 & -1 & 0
\\\end{array}\right) \left(\begin{array}{ccc} j_1 & j_2 & j_3 \\
-m_1 & m_2 & m_3 \\\end{array}\right)\end{eqnarray} Eq(\ref{3j0})
may be derived from Eq(\ref{C}) by using the identity
$\sqrt{j(j+1)}\vec{Y}_{jm}^{(0)}=\vec{L}Y_{jm}$ (where $\vec{L}$
is the angular momentum operator) and integration by parts. The
relation(\ref{3j1}) was found with the help of Eq. (18) in Ref.
\cite{Dowker}.

\subsection{Spherical scatterers}
Assuming spherically symmetric scatterers, one may define phase shifts
$\delta_{TE}^j(\omega),\delta_{TM}^j(\omega)$ (by parity these two channels do not mix).
Similarly to the scalar case we use the notation
$t_{j\alpha}={1\over2}(-1)^{j+\alpha}(e^{2i\delta_j^{(\alpha)}(i\omega)}-1)$.

Choosing the $z$-axis along $\hat{a}$ the operator $K=TGTG$ splits
to independent blocks $K^{(m)}$ corresponding to the values of the
azimuthal number $m$. In a given block the $g$-matrix elements
become
\begin{eqnarray}&
g^{(m)}_{j\alpha;j'\alpha'}=\\
\nonumber & \sum_{l}\sqrt{{8\over\omega a}(2l+1)}
K_{l+{1\over2}}(\omega a)\int d\Omega Y_{l,0}
\left(\vec{Y}_{jm}^{(\alpha)*}\cdot\vec{Y}_{j'm}^{(\alpha')}\right)\end{eqnarray}
The matrix $K^{(m)}(i\omega)$ is then written explicitly as:
$$K^{(m)}_{j\alpha;j''\alpha''}(i\omega)=
t^{(A)}_{j\alpha}g^{(m)}_{j\alpha;j'\alpha'}t^{(B)}_{j'\alpha'}g^{(m)}_{j'\alpha';j''\alpha''}$$

\begin{figure}[tbh]
\includegraphics[scale=0.8]{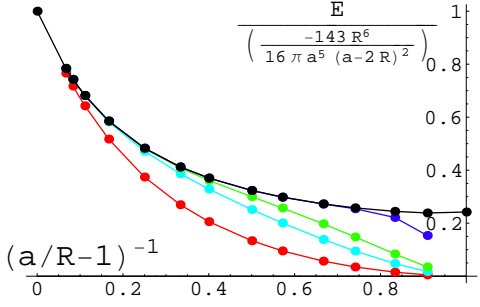}
\caption{EM Casimir energy of two conducting spheres of radius $R$ at
distance $a$ between their centers.
The graphs show $E/E_0$ as a function of $(a/R-1)^{-1}$ where
$E_0$ is the large distance asymptotic expression of it.
$E_0^S=-{143R^6\over16\pi(a-2R)^2a^5}$.
The black curve shows the calculated exact result for
$j_0\rightarrow\infty$. We extrapolated it to $a=2R$ and
$a=\infty$ using the known asymptotics. The colored graphs show
the result of including partial waves of $j\leq j_0$ where:
$j_0=1$ (red), $j_0=2$ (sky blue),
$j_0=4$ (green), $j_0=10$ (blue).}
\label{Eelm}
\end{figure}

In the particular case of a perfectly conducting sphere of radius $R$ one has:
\begin{eqnarray}& t_{TE}^j(i\omega)={\pi\over2}{I_{j+{1\over2}}
(\omega R)\over K_{j+{1\over2}}(\omega R)},\\ &
t_{TM}^j(i\omega)=-{\pi\over2}{{d\over
dx}\left(\sqrt{x}I_{j+{1\over2}}(x)\right)\over{d\over dx}
{\left(\sqrt{x}K_{j+{1\over2}}(x)\right)}}\Big|_{x=\omega
R}.\end{eqnarray}
Using this we numerically calculated the
electromagnetic Casimir energy for a pair of conducting spheres at
distance $a$ between their centers. As in the scalar case writing
$K=\tilde{K}^2$ and considering $\det(1\pm\tilde{K})$ separately
allowed us to also find the interaction energies $E_e,E_m$ of a sphere
near a conducting/infinitely permeable mirror placed $a/2$ from its center.
The two spheres energy is then the sum $E_s=E_e+E_m$.
Most of the calculations where done by including modes having $j\leq 10$,
however for the shortest distances $a=2.35,2.2,2.1$ were
convergence is slower, we extended the retained modes up to
$j=20,40,60$ respectively. The results are shown in the following
table (\ref{TableEM}) (written in units where $R=1$) and in Fig. \ref{Eelm}

\begin{eqnarray}
\begin{tabular}{|l|l|l|l|}
\hline
$a$&$ E_e$ &$E_m$&$E_s$\\
\hline
2.1 & $-16.15   $ & 14.5   &      $-1.662    $         \\
2.2 & $ -3.82   $ & 3.48   &      $-0.337635 $         \\
2.35& $ -1.157  $ & 1.073  &      $-8.356 \cdot10^{-2}$\\
2.5 & $ -0.53   $ & 0.50   &      $-3.18\;\cdot10^{-2}$\\
2.75& $ -0.211  $ & 0.201  &      $-9.595\cdot10^{-3}$ \\
3   & $ -0.1074 $ & 0.1036 &      $-3.787\cdot10^{-3}$ \\
3.5& $-3.97    \cdot10^{-2}$ & $3.88  \cdot10^{-2}$ & $-8.917 \cdot10^{-4}$\\
4 &  $-1.89    \cdot10^{-2}$ & $1.86  \cdot10^{-2}$ & $-2.864 \cdot10^{-4}$\\
5 &  $-6.24    \cdot10^{-3}$ & $6.19  \cdot10^{-3}$ & $-4.887 \cdot10^{-5}$\\
7 &  $-1.38    \cdot10^{-3}$ & $1.37  \cdot10^{-3}$ & $-3.965 \cdot10^{-6}$\\
10&  $-3.06    \cdot10^{-4}$ & $3.06  \cdot10^{-4}$ & $-3.032 \cdot10^{-7}$\\
13&  $-1.04    \cdot10^{-4}$ & $1.04  \cdot10^{-4}$ & $-4.703 \cdot10^{-8}$\\
16&  $-4.47    \cdot10^{-5}$ & $4.47  \cdot10^{-5}$ & $-1.085 \cdot10^{-8}$\\
\hline
\end{tabular}
\label{TableEM}
\end{eqnarray}

\begin{figure}[tbh]
\includegraphics[scale=0.8]{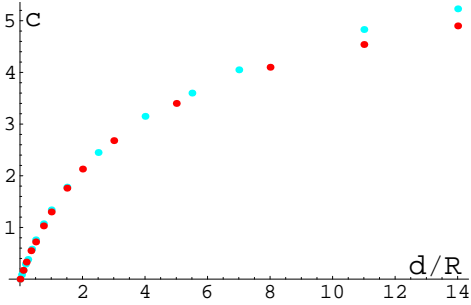}
\caption{Including partial waves of $j\leq j_0$ results in error
behaving roughly as $e^{-cj_0}$. The graph shows the constant $c$
as a function of the separation distance $d$. The blue dots
correspond to our results for two conducting spheres (where
$d=a-2R$) and the red dots to conducting-sphere + conducting plate
(where $d=a/2-R$). At small distances both cases give $c\sim
1.7d/R$.} \label{c}
\end{figure}

The numerical results seem to converge as $j_0\rightarrow\infty$
at roughly an exponential rate. The graph in Fig. \ref{c}
shows how the speed of convergence depends on the distance between
the bodies. It is interesting to note that the results obtained in
section \ref{sswe}  for the scalar case give almost the same
graph. Also, one can easily check that the results for $E_s$ are
basically the same as the ones obtained in \cite{EmigGraham}
taking into account we chose to normalize the energy in comparison
with the large distance asymptotic expression for the energy.

\appendix
\numberwithin{equation}{section}

\section{Proof of the Green's function expansions eqns
(\ref{YCY},\ref{D_0expansion})}\label{AGB}
\subsection{Scalar case }
Suppose $\vec{R}=\vec{a}+\vec{r}'$ then obviously
$e^{i\vec{k}\cdot\vec{R}}=e^{i\vec{k}\cdot\vec{a}}e^{i\vec{k}\cdot\vec{r}'}$.
Inserting the well known expansion
$$e^{i\vec{k}\cdot\vec{r}}=4\pi\sum
i^lY_{lm}^*(\hat{k})Y_{lm}(\hat{r})j_l(kr)$$
We get
$$\sum i^lY_{lm}^*(\hat{k})Y_{lm}(\hat{R})j_l(kR)=$$
$$4\pi\left(\sum i^{l''}Y_{l''m''}^*(\hat{k})Y_{l''m''}(\hat{a})j_{l''}(ka)\right)$$
$$\times\left(\sum i^{l'}Y_{l'm'}^*(\hat{k})Y_{l'm'}(\hat{r}')j_{l'}(kr')\right)$$
Multiplying both sides by $Y_{lm}(\hat{k})$ and integrating $\int d\Omega_k$
we find
\begin{eqnarray}\label{jys}&
j_l(kR)Y_{lm}(\hat{R})=\\
\nonumber & 4\pi\sum_{l'm'l''m''}\left(\int d\Omega
Y_{lm}Y_{l'm'}^*Y_{l''m''}^*\right)\\   \nonumber & \times
i^{l''+l'-l}
j_{l''}(ka)Y_{l''m''}(\hat{a})j_{l'}(kr')Y_{l'm'}(\hat{r}')
\end{eqnarray}
Concentrating on the case $R,a>r'$, it makes sense to separate the ingoing and
outgoing parts in the last equation.
This amounts to replacing the bessel functions $j_l(kR),j_l(ka)$ by hankel functions
$h_l(kR),h_l(ka)$ of the first or second type corresponding to outgoing or ingoing waves.
Since this argument may seem as hand-waving, we will return and elaborate
on it more at the end of the proof.
Equating the outgoing parts we have:
\begin{eqnarray}\label{hys}&
h_l^{(1)}(kR)Y_{lm}(\hat{R})=\\   \nonumber &
4\pi\sum_{l'm'l''m''}
\left(\int d\Omega Y_{lm}Y_{l'm'}^*Y_{l''m''}^*\right) \\
\nonumber &\times  i^{l''+l'-l}
h^{(1)}_{l''}(ka)Y_{l''m''}(\hat{a})j_{l'}(kr')Y_{l'm'}(\hat{r}')
\end{eqnarray}

It is well known that for $R>r$ the free propagator may be expanded as
$$-{1\over4\pi}{1\over |\vec{R}-\vec{r}|}e^{ik|\vec{R}-\vec{r}|}=
-ik\sum j_l(kr)h_l^{(1)}(kR)Y_{lm}^*(\hat{r})Y_{lm}(\hat{R})$$
Substituting here Eq. (\ref{hys}) we finally get 
\begin{eqnarray}& G(\vec{r},\vec{a}+\vec{r}')=\\   \nonumber &
-4\pi i\omega\sum i^{l''+l'-l} \left(\int d\Omega
Y_{lm}Y^*_{l'm'}Y^*_{l''m''}\right)
\\ \nonumber & \times j_l(\omega r)j_{l'}(\omega r')h^{(1)}_{l''}(\omega a)
Y_{lm}^*(\hat{r})Y_{l'm'}(\hat{r}')Y_{l''m''}(\hat{a})\end{eqnarray}
Which is exactly Eq. (\ref{YCY}).


Let us now return to the derivation of Eq. (\ref{hys}) from Eq.
(\ref{jys}). We first note that the function
$h_{l_0}^{(1)}(kR)Y_{l_0m_0}(\hat{R})$ with
${\vec{R}=\vec{a}+\vec{r}}$ being a  solution of the free wave
equation may be expanded around $\vec{r}=0$ in the form
$$h_{l_0}^{(1)}(kR)Y_{l_0m_0}(\hat{R})=\sum\left(c_{lm}^{(1)} h^{(1)}_l(kr)
+\tilde{c}_{lm}^{(1)} h^{(2)}_l(kr)\right) Y_{lm}(\hat{r})$$
for some ($\vec{a}$ dependent) constants $c_{lm}^{(1)},\tilde{c}_{lm}^{(1)}$.
To be more precise $h_{l_0}^{(1)}(kR)Y_{l_0m_0}(\hat{R})$ is a solution only for
$\vec{r}\neq-\vec{a}$ (i.e. $\vec{R}\neq0$) therefore one has two separate expansions:
one for $r<a$ and another for $r>a$. We concentrate on the latter.

Since $h_{l_0}^{(1)}(kR)Y_{l_0m_0}(\hat{R})$ is a purely outgoing wave it is clear that
the expansion in terms of $\vec{r}$ must also contain only outgoing waves i.e. $\tilde{c}_{lm}^{(1)}\equiv0$.
This claim is based on ``physical intuition''. A more rigorous mathematical argument
may be constructed by considering first pure imaginary $k=iq$ with $q>0$. One then note that
$h_{l_0}^{(1)}(iqR)$ is exponentially decreasing as $R\rightarrow\infty$ which imply that
the same must hold for the r.h.s. Since the $Y_{lm}$'s are linearly independent this require
all the $\tilde{c}_{lm}^{(1)}$'s to vanish.

A similar expansion obviously exists also for $h^{(2)}$:
$$h_{l_0}^{(2)}(kR)Y_{l_0m_0}(\hat{R})=\sum c_{lm}^{(2)} h^{(2)}_l(kr)Y_{lm}(\hat{r})$$
Summing the two expansions we have
$$j_{l_0}(kR)Y_{l_0m_0}(\hat{R})\equiv{1\over2}\left(h^{(1)}_{l_0}(kR)+h_{l_0}^{(2)}(kR)\right)Y_{l_0m_0}(\hat{R})=$$
$$={1\over2}\sum \left(c_{lm}^{(1)} h^{(1)}_l(kr)+c_{lm}^{(2)} h^{(2)}_l(kr)\right)Y_{lm}(\hat{r})$$
However such an expansion is clearly unique. Therefore it must be
the same as the expansion in Eq. (\ref{jys}). Comparing the two
(and using $j_l\equiv {1\over2}(h^{(1)}+h^{(2)})$) we deduce
$$c^{(1)}_{lm}=c^{(2)}_{lm}=$$
$$4\pi\sum_{l''m''} i^{l''+l-l_0}\left(\int d\Omega Y_{l_0m_0}Y_{lm}^*Y_{l''m''}^*\right)
\times j_{l''}(ka)Y_{l''m''}(\hat{a})$$ which proves Eq.
(\ref{hys}).

\subsection{The electromagnetic case }
To derive the EM expansion (\ref{D_0expansion}) we similarly start by using the identity
\begin{eqnarray}&{e^{i\vec{k}\cdot\vec{r}}}\times\tens{1}=\\ \nonumber & 4\pi\sum
i^lj_l(kr)\vec{Y}_{jlm}^*(\hat{k})\otimes\vec{Y}_{jlm}(\hat{r})\end{eqnarray}
Repeating the same steps as for the scalar we then find that
$$\tens{G}_0=-{e^{i\omega r}\over4\pi r}\times\tens{1}$$
may be expanded as
\begin{eqnarray}\label{G0te}
\tens{G}_{\omega}=  |(j l m)_B\ra{{\cal C}_{jlm;j'l'm'}}\la(j'l'm')_A|
\end{eqnarray}
where
\begin{eqnarray}\label{vyj}
|(j l m)_{A,B}\ra=\sqrt{2\omega^2\over\pi} j_l(\omega
r_{A,B})\vec{Y}_{jlm}(\hat{r}_{A,B})
\end{eqnarray}
are the free vectorial spherical wave functions centered at $P_A,P_B$.
The $\cal C$ coefficients may be written as
\begin{eqnarray}&
{\cal C}_{jlm;j'l'm'}=-{i\pi\over 2\omega} \sum_{l'',m''} \Big[\\
\nonumber &
\tilde{C}_{jj'}\left(\begin{array}{ccc} l & l' & l'' \\
m & m' & m''\\\end{array}\right) i^{l''+l'-l} h^{(1)}_{l''}(\omega
a) Y_{l''m''}(\hat{a})\Big], \end{eqnarray} Here $\vec{Y}_{jlm}$
are vectorial spherical harmonics, $Y_{lm}$ are the usual scalar
spherical harmonics, and $j_l,h_l$ are spherical Bessel and Hankel
functions. The coefficients
$$\tilde{C}_{jj'}\left(\begin{array}{ccc} l & l' & l'' \\
m & m' & m'' \\\end{array}\right)$$ are found to be expressed as
the following integral of spherical functions:
\begin{eqnarray}&
\tilde{C}_{jj'}\left(\begin{array}{ccc} l & l' & l'' \\  m & m' &
m'' \\\end{array}\right)= \\ \nonumber & 4\pi \int d\Omega
(\vec{Y}_{jlm}\cdot\vec{Y}_{j'l'm'}^*)Y_{l''m''}^*\end{eqnarray}

The radiation gauge propagator ${\cal D}_0$
is given by the transverse part of $\tens{G}_0$.
In Eq(\ref{vyj}) each $j,m$ correspond to three different spherical function
$|j l m\ra$ (having $l=j-1,j,j+1$).
These may be decomposed in terms of the
{\bf TE} and {\bf TM} modes and a nonphysical longitudinal mode.
$$|TE\ra=|j j m\ra$$
$$|TM\ra=\sqrt{j+1\over 2j+1}|j,j-1, m\ra-\sqrt{j\over 2j+1}|j,j+1, m\ra$$
$$|L\ra=\sqrt{j\over 2j+1}|j,j-1, m\ra+\sqrt{j+1\over 2j+1}|j,j+1, m\ra$$

To obtain the required expansion of the radiation gauge propagator ${\cal D}_0$
we need to rewrite Eq(\ref{G0te}) in terms of these three modes and drop the
parts containing the longitudinal mode.
This can be done quite straightforwardly leading to the results (\ref{D_0expansion}-\ref{Yvec}).

\section{Analytical properties of the
$T_AG_0T_BG_0$-operator:}\label{mathp}

Having established the form \eqref{F for bulk} for the energy, we
turn here to discuss the properties of this expression.
The main aim of this appendix is to rigorously show that the object
$\log\det(1-T_A{G_0}_{AB}T_B{G_0}_{BA})$ is well defined and finite.
The main mathematical notions and theorems which we use here,
are briefly reviewed in appendix \ref{ccm}.

As already remarked in the introduction it is well known that
$\det(1-M)$ is well defined whenever $M$ is a {\it trace class}(t.c.)
operator (definition \ref{trace class definition}).
We would like to show that for a large class of situations
(including a pair of disjoint finite bodies $A,B$, separated by a
finite distance) the operator
$T_A{G_0}_{AB}T_B{G_0}_{BA}:H_A\rightarrow H_A$ is trace class in
the continuum limit , and so prove that indeed the expression
\eqref{F for bulk} is finite and well defined.

Indeed, by theorem \ref{sb} the mere fact that $G_0(x,y)$ is a
smooth function for $x\neq y$ is sufficient to guarantee that for
any pair of compact volumes $A,B\in\mathbb{R}^3$ at finite mutual
distance the operator ${G_0}_{AB}$ is trace class. To deduce that
$T_A{G_0}_{AB}T_B{G_0}_{BA}$ is trace class (and by similar
argument also $1-{G_0}{\cal J}T_A$ appearing in Eq. (\ref{AD})) it
is then enough (proposition \ref{tb}) to make sure
$T_{A,B}(i\omega)$ are bounded (definition \ref{dbo}).

In the context of dielectric interaction, it is particularly easy
to show that $T(i\omega)$ is bounded. 
In physical systems at equilibrium, it follows from causality
properties of the dielectric function \cite{LifsitzPitaevskii},
that $\chi(i\omega,x)\geq0$.  We then have the following
\begin{lem}\label{T<}
For $\chi(i\omega,x)>0$ , the $T$ operators are positive and
bounded.
\end{lem}
{\it Proof:} Since $G_0,\chi>0$ (definition \ref{positive op}) one
may write $T=\sqrt{\chi}{\omega^2\over
1+\omega^2\sqrt{\chi}G_0\sqrt{\chi}}\sqrt{\chi}$ from which it is
seen that $T>0$ and that in the operator norm
$||T||\leq\omega^2||\chi||$ $\square$.\\
In fact, this holds also for nonlocal $\chi$ as long as
$f(x)\mapsto \int_{A}\chi(i\omega,x,x')f(x')\dd x'$ is a
bounded positive operator $H_A\rightarrow H_A$. In the context of
more general type of interactions which may not be positive, one
needs to use some assumption on the stability of the system to
guarantee that $T(i\omega)$ is bounded. We do not elaborate on
this here.

An alternative approach to proving the trace class property of
$T_A{G_0}_{AB}T_B{G_0}_{BA}$ is based on the notion of a
Hilbert-Schmidt operator (definition \ref{HS}, also denoted
$H.S.$). Here the frequently used strategy in operator analysis is
to use the following fact: if $U\in H.S.$ and $V\in H.S.$, then
$UV\in t.c.$. The advantage of this approach is that it is very
easy to check if an operator is Hilbert \-Schmidt. Since the
Hilbert Schmidt norm is $||A||_{H.S.}^2=\Tr (A^{\dag}A)$, one may
evaluate it directly, (e.g. by computing $\int |A(x,x')|^2$).
\begin{thm} For any two bodies $A$, $B$ such that $\int_{A\times
B}\dd x\dd y|{G_0}(x,y)|^2<\infty $, $T_A{G_0}_{AB}T_B{G_0}_{BA}$
is trace class.
\end{thm}
{\it Proof}: First we show that $T_A{G_0}_{AB}$ and
$T_B{G_0}_{BA}$ are Hilbert Schmidt operators. This can be
verified in the following way. We have just seen that $T_A,T_B$
are bounded operators. Now note that ${G_0}_{AB}$ is Hilbert-Schmidt,
since,
\begin{eqnarray}
||{G_0}_{AB}||_{H.S}^2=\int_{A\times B}\dd x\dd
y|{G_0}_{AB}(x,y)|^2,
\end{eqnarray}
which is finite under the condition above. Now the inequality
$||T_A{G_0}_{AB}||_{H.S}\leq ||T_A||||{G_0}_{AB}||_{H.S}$ implies
that $T_A{G_0}_{AB}$ is Hilbert Schmidt. Finally using $U,V\in
H.S.\Rightarrow UV\in t.c.$ we see that
$T_A{G_0}_{AB}T_B{G_0}_{BA}\in t.c. $ $\square$

\begin{cor}
For any finite bodies $A,B$, such that $distance(A,B)>0$, and any
Green's function which is finite away from the diagonal,
$T_AG_0T_BG_0\in t.c.$
\end{cor}
\begin{exa}
For the scalar field discussed above,
$G_0(x,y)={e^{-\omega|x-y|}\over4\pi |x-y|}$, the condition is
satisfied. In the same way it is satisfied for the electromagnetic
field (one has to take into account also matrix indices but these
discrete indices do not change finiteness of the integrals)
\end{exa}

\begin{rem}
The $\omega$ integration in Eq. \eqref{F for bulk}, is convergent.
To see this note that $G_0$ decays exponentially with $\omega$
therefore, $||G_0||_{H.S.}$ decays exponentially, also the
$||T||$'s do not grow more then quadratically in $\omega$.

In the EM case one may also worry due to the factor
${1\over\omega^2}$ appearing in ${{\cal
D}_0}_{ij}(x,y)=(\delta_{i,j}-{1\over\omega^2}\nabla_i^{(x)}\nabla_j^{(y)})G_0(x,y)$,
about convergence for $\omega\sim0$. This factor however gets
cancelled since $||T||\leq\omega^2||\chi||$ as shown in lemma
\ref{T<}.

\end{rem}

\begin{figure}
\includegraphics[scale=0.4]{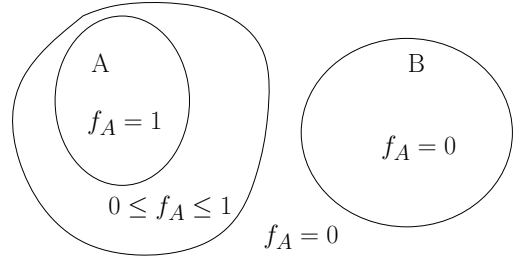}\caption{The support of the function $f_A$}
\label{supportF}
\end{figure}

One may also show that ${G_0}_{AB}$ are t.c. themselves by using
H.S. properties. The bodies are assumed not to touch, thus we can
choose a $C^{\infty}_0$ (compactly supported and infinitely
smooth) function $f_A$, such that $P_Af_A=P_A$, and $P_Bf_A=0$
where $P_A,P_B$ are the projections on $L^2(A),L^2(B)$ (i.e.
$f_A(x)=1$ for $x\in A$, and it then smoothly goes to $0$, before
reaching body $B$ see Fig.\ref{supportF}) .

Writing:
\begin{eqnarray}&
G_{0AB}=L_1L_2\\ \nonumber & L_1=P_A{1\over(
p^2+\omega^2)^\alpha}\,\,\,;\,\,\,L_2=(p^2+\omega^2)^{\alpha}
f_AG_0P_B,
\end{eqnarray}
we see that if $4\alpha>d$,
\begin{eqnarray}&
||L_1||_{H.S.}^2=\Tr(P_A{1\over (p^2+\omega^2)^{\alpha}})
(P_A{1\over (p^2+\omega^2)^{\alpha}})^\dag=\\
\nonumber &Vol(A)\int\D^d p |{1\over
(p^2+\omega^2)^{2\alpha}}|<\infty
\end{eqnarray}
and so $L_1$ is Hilbert Schmidt. Next, we check that $L_2\in
H.S.$. To see this last point, note that
\begin{eqnarray}&
<x|L_2|x'>=<x|(p^2+\omega^2)^{\alpha} f_AG_0P_B|x'>= \\ \nonumber
& (-\triangle_x+\omega^2)^{\alpha} f_A(x)G_0(x-x')P_B(x')
\end{eqnarray}
Since $G_0(x-x')$ is smooth away from $x=x'$, where the expression
is anyway zero because $f_AP_B=0$, and since $\la x|L_2|x'\ra$ has
compact support (for integer $\alpha$) we see that $||L_2||_{H.S.}^2= \int \D x\D
x'|L_2|^2<\infty$. Thus, $G_{0AB}$ can be written as a product of
two H.S. operators, and as such is trace class.

Finally, we have that
\begin{thm}\label{g<1} ({\bf Eigenvalues of TGTG}) For $\chi>0$, {\it all} eigenvalues
$\lambda$ of the (compact) operator $T_A{G_0}_{AB}T_B{G_0}_{BA}$
appearing in \eqref{F for bulk} satisfy $1>\lambda\geq 0$.
\end{thm}
{\it Proof:} We will use repeatedly that for bounded operators
$X,Y$ the nonzero eigenvalues of $XY$ and $YX$ are the same. Note
first that $G_0,\chi\geq0$ (as operators) implies
\begin{eqnarray*}{\rm spec}(\chi G_0)\setminus\{0\}={\rm
spec}(\sqrt{G_0}\chi\sqrt{G_0})\setminus\{0\}\subset [0,\infty).
\end{eqnarray*}
Writing $T_\alpha G_0=1-{1\over 1+\omega^2\chi_{\alpha} {G_0}}$ as
an operator on $L^2(\R^3)$ it is then clear that its spectrum lies
in [0,1). The same conclusion then applies to the operator
$\sqrt{G_0}T_\alpha\sqrt{G_0}$ but since it is hermitian one
concludes also $||\sqrt{G_0}T_\alpha\sqrt{G_0}||<1$ from which it
follows $||\sqrt{G_0}T_A G_0 T_B\sqrt{G_0}||<1$ and hence
$\lambda<1$. Similarly $\sqrt{G_0}T_\alpha\sqrt{G_0}\geq 0$ imply
$\lambda\geq 0$ $\square$

\section{Some properties of (infinite dimensional)operators}\label{ccm}
Here we recall some mathematical notions that we have used in
describing the trace class properties of Eq. \eqref{F for bulk}.
\begin{definition}
For an operator $B:H\rightarrow H$, the operator norm of $||B||$
is defined as $||B||={\rm sup}_{\psi\in H, \psi\neq 0}
{|<\psi|B|\psi>|\over <\psi|\psi>}$
\end{definition}
\begin{definition}\label{dbo}
An operator $B$ is bounded if $||B||<\infty$
\end{definition}

\begin{definition}\label{positive op}
An operator $A:H\rightarrow H$ is called a {\bf positive operator}
(denoted $A>0$) iff $\langle\psi|A|\psi\rangle\geq 0$ for every
$\psi\in H$.
\end{definition}
This implies that $A$ is hermitian and its spectrum nonnegative.
If $A:H\rightarrow H$ is a positive operator then there exist a
unique positive operator $B:H\rightarrow H$ satisfying $A=B^2$. B
is called the {\bf square root} of $A$ and denoted $\sqrt{A}$.


\begin{definition}\label{trace class definition}
An operator  $A:H_1\rightarrow H_2$ is called {\bf trace class}
(and denoted $A\in t.c.$ or $A\in{\cal J}_1$) iff
$\sum||A\psi_n||<\infty$ where $\{\psi_n\}_{n=1}^\infty$ is some
orthonormal basis of $H_1$. It can be shown that this condition
does not depend on the choice of the orthonormal basis. (Note that
the definition makes sense even when $H_1\neq H_2$.)
\end{definition}

If $A:H\rightarrow H$ is trace class then for any orthonormal basis
$\{\psi_n\}_{n=1}^\infty$ of $H$ the sum
$\sum\langle\psi_n|A|\psi_n\rangle$ converges to the same (finite)
value which is denoted $tr(A)$ and called the trace of $A$. One
then also have $tr(A)=\sum\lambda_n$ where $\{\lambda_n\}$ are the
eigenvalues of $A$ (Lidski's theorem)

If $A:H\rightarrow H$ is trace class then the determinant
$\det(1+A)$ may also be rigorously defined and one has
$\det(1+A)=\prod(1+\lambda_n)$.

The following theorem may be proved
using the well known fact that the Fourier coefficients of a smooth $K(x,y)$
decay faster then any power. (Note that these coefficient also serve as the matrix elements
with respect to Fourier basis of the operator defined by $K$.)
\begin{thm}\label{sb}
Consider an operator $A:L^2(D_1)\rightarrow L^2(D_2)$ where
$D_1,D_2$ are some domains in $\R^n$ which is given explicitly as
an integral $A\psi(x)=\int_{D_1}K(x,y)\psi(y)dy$. A sufficient
condition for $A$ to be trace class is that $D_1,D_2$ are compact
and $K(x,y)$ is smooth in a neighborhood of $D_1\times D_2$.
\end{thm}

\begin{prop}\label{tb}
If $A$ is trace class and $B$ bounded then $AB$ and $BA$ are also
trace class and $Tr(|AB|),Tr(|BA|)\leq ||B||Tr(|A|)$.
\end{prop}

\begin{definition}\label{HS}
$M$ is a Hilbert Schmidt operator (denoted $M\in H.S.$ or $M\in J_2$) if
$||M||^2_{H.S.}\equiv\Tr M^{\dag}M<\infty$
\end{definition}
In particular we mention that the product of two Hilbert Schmidt operators
always give a trace class operator.


\begin{thebibliography}{00}




%
\bibitem{Casimir48} H. B. G. Casimir, Proc. Koninkl. Ned. Akad. Wet. {\bf 51},
793 (1948).
%
\bibitem{Lamoreaux97}
S. K. Lamoreaux, Phys Rev. Lett. {\bf 78} 5-8(1997);
%

%
\bibitem{MohideenRoy98}
U. Mohideen and A. Roy  Phys. Rev. Lett. {\bf 81} 4549 (1998).
%
\bibitem{Bressi}
G. Bressi, G. Carugno, R. Onofrio and G. Ruoso, Phys. Rev. Lett.
{\bf 88}, 041804 (2002).
%

\bibitem{Geyer}
B. Geyer, G. L. Klimchitskaya, and V. M. Mostepanenko, Phys. Rev.
A {\bf 67}, 062102 (2003).

\bibitem{Pirozhenko Lambrecht Svetovoy06}
I. Pirozhenko, A. Lambrecht, V. B. Svetovoy, New Journal of
Physics {\bf 8}, 238 (2006)
%
\bibitem{KardarGolestanian}
M. Kardar and R. Golestanian, Rev. Mod. Phys. {\bf 71}, 1233
(1999).
%
\bibitem{BordagMohideenMostepanenko}
M. Bordag M., U. Mohideen and V.M. Mostepanenko,
Phys.Rept.{\bf 353}:1-205,(2001).

\bibitem{Miltons Book}
K A Milton, {\it The Casimir Effect: Physical Manifestations of
Zero-Point Energy}, World Scientific, 2001.

\bibitem{Millonis Book}
P. W. Miloni, {\it The Quantum Vacuum, An Introduction to Quantum
Electrodynamics}, (Academic Press, San Diego 1994).

\bibitem{Ashourvan Miri Golestanian}
A. Ashourvan, M. Miri, R. Golestanian, Phys. Rev. Lett. {\bf 98}, 140801
(2007)

\bibitem{Emig07}
T. Emig, Phys. Rev. Lett. {\bf 98},160801 (2007).

\bibitem{Gies}
H. Gies and K. Klingmuller, Phys.Rev.Lett.{\bf 97}, 220405 (2006);
Phys.Rev. D{\bf 74},  045002(2006).

\bibitem{RodriguezIbanescuIannuzzi}
A. Rodriguez, M. Ibanescu, D. Iannuzzi, J. D. Joannopoulos, S. G.
Johnson, Phys. Rev. A, {\bf 76}, 032106 (2007). preprint arXiv:0705.3661

\bibitem{KennethKlich06}
O. Kenneth and I. Klich, Phys. Rev. Lett. {\bf 97}, 160401 (2006).

\bibitem{Lifshitz}
E. M. Lifshitz, Sov. Phys. JETP {\bf 2}, 73 (1956).
%
\bibitem{BalianDuplantier78}
R. Balian and B. Duplantier, Ann. Phys.(N.Y.) {\bf 112}, 165, (1978).
%

\bibitem{Reed Simon Scattering}
M. Reed and  B. Simon, {it Methods of mathematical physics 3:
Scattering Theory}, Academic Press 1979.

\bibitem{Bachas}
C P Bachas  2007 J. Phys. A: Math. Theor. {\bf 40} 9089(2007).
quant-ph/0611082
\bibitem{ZNussinov}
Z. Nussinov, cond-mat/0107339 (Appendix A and footnote [35]
therein).
\bibitem{EmigGraham}
T. Emig, N. Graham, R. L. Jaffe and M. Kardar,
Phys.Rev.Lett.{\bf 99},170403 (2007). arXiv:0707.1862v2
%
\bibitem{Kenneth99}
O. Kenneth, preprint hep-th/9912102.

\bibitem{Genet Lambrecht Renaud}
C. Genet, A. Lambrecht and S. Reynaud, Phys. Rev. A{\bf 67},
043811 (2003);
M. T. Jaekel and S. Reynaud, J. Phys. I {\bf 1}, 1395 (1991).
quant-ph/0101067.

\bibitem{BulgacMagierskiWirzba}
A. Bulgac, P. Magierski and A. Wirzba, Phys. Rev. D{\bf 73} ,025007 (2006).

\bibitem{Wirzba97}
A. Wirzba, Phys. Rept. {\bf 309}, 1-116 (1999).
%
\bibitem{LiKardar}
H. Li and M. Kardar, Phys. Rev. Lett. {\bf 67}, 3275 (1991); H. Li and
M. Kardar, Phys. Rev. A{\bf 46}, 6490 (1992).
%
\bibitem{FeinbergMannRevzen}
J. Feinberg, A. Mann and M. Revzen, Annals of Physics (New York)
{\bf 288}, 103-136 (2001).
%
\bibitem{Simon79}
B. Simon. {\it Trace ideals and their applications}, LMS vol 35.
Cambridge University Press, New York, 1979.

\bibitem{Candelas Deutsch}
P. Candelas and D. Deutsch,  Phys. Rev. D{\bf 20} ,3063 (1979).

\bibitem{Buscher Emig} R. Büscher, T. Emig
Phys. Rev. A{\bf 69}, 062101 (2004).
%
\bibitem{LifsitzPitaevskii}
E. M. Lifshitz and L. P. Pitaevskii, Statistical Physics, Pt. 2,
Pergamon, Oxford, 1984.
%
\bibitem{kn}
O. Kenneth and S.Nussinov, Phys. Rev. D{\bf 65}, 085014 (2002).
\bibitem{Kampen}
N. G. van Kampen, B. R. A. Nijboer and K. Schram, Phys. Lett. {\bf 26}A,
7, 307 (1968).

\bibitem{Mello Stone}
P. A. Mello and D. Stone, Phys. Rev. B{\bf 44}, 3559 (1991).

\bibitem{Dowker}
J.S. Dowker Class. Quantum Grav. {\bf 7}, 1241-1251 (1990).

\end{thebibliography}
\end{document}